\documentclass[12pt,draftclsnofoot,peerreview,oneside,onecolumn]{IEEEtran}

\usepackage{amssymb}
\usepackage{pstricks}
\usepackage{amsmath}
\usepackage[dvips]{graphicx}
\usepackage{colortbl}
\usepackage{enumerate}
\usepackage{ifthen}
\newtheorem{theorem}{Theorem}
\newtheorem{corollary}{Corollary}

\newtheorem{lemma}{Lemma}

\begin{document}

\title{Performance of Multiantenna Linear MMSE Receivers in Doubly Stochastic Networks}
\author{Junjie~Zhu,~\IEEEmembership{Student Member},
  Siddhartan~Govindasamy,~\IEEEmembership{Member}, Jeff Hwang\thanks{J. Zhu and S. Govindasamy are with Franklin W. Olin College of Engineering, Needham, MA. Jeff Hwang was with Olin College and is currently with Intel Corp. and Stanford University. email: junjie.zhu@students.olin.edu, siddhartan.govindasamy@olin.edu (corresponding author), jeff.hwang@alumni.olin.edu.
  Portions of this material have been presented at the 2012 and 2013 IEEE International Conference on Communications (ICC). This research was supported in part by the National Science Foundation under Grant CCF-1117218.}}
  \maketitle

\pagenumbering{arabic}
\begin{abstract}
A technique is presented to characterize the
Signal-to-Interference-plus-Noise Ratio (SINR) of a representative link with a multiantenna linear Minimum-Mean-Square-Error receiver in a wireless network with transmitting nodes distributed according to a  doubly stochastic process, which is a generalization of the Poisson point process. The cumulative distribution function of the SINR of the representative link is derived  assuming independent Rayleigh fading between antennas.
Several representative spatial node distributions are considered, including
networks with both deterministic and random clusters, strip networks (used to model roadways, e.g.), hard-core networks and networks with generalized path-loss models.
In addition, it is shown that if the number of antennas at the
representative  receiver is increased linearly with the nominal
node density, the signal-to-interference ratio converges in distribution to  a random variable that is non-zero in general, and  a positive constant in certain cases. This result indicates that to the extent that the
system assumptions hold, it is possible to scale such networks
by increasing the number of receiver antennas linearly with the
node density. The results presented here are useful in
characterizing the performance of multiantenna wireless
networks in more general network models than what is currently available.
\end{abstract}


\section{Introduction}

Multiantenna systems can  increase data rates
in  wireless networks through spatial multiplexing, beamforming
and interference mitigation, the performance of which is
highly dependent on the spatial separations between nodes. Most of the results in the literature that explicitly
model multiantenna systems in spatially distributed networks have
focused on homogenous Poisson spatial node distributions, i.e.
systems where node positions are independent of one another and
are distributed uniformly randomly on a plane (\cite{govindasamy2007spectral},
\cite{jindal2011multi}, \cite{ali2010performance},
\cite{louie2011open}). While simpler and more tractable,
homogenous Poisson spatial node distributions may not apply in many
scenarios, such as networks with hot spots, clusters of active nodes or restrictions on the locations of nodes (such as vehicular networks where nodes are restricted to being on a roadway).  In particular, networks where active nodes are spatially correlated, such as doubly stochastic networks, are  difficult to analyze. Giacomelli, Ganti and Haenggi  remark in \cite{giacomelli2011outage} that ``Extensions [of results for homogenous PPP networks] to models with dependence (node repulsion or attraction) are non-trivial." In this work, we provide exact (and in some cases, closed-form) expressions for the CDF of the SINR for several different clustered network models which are examples of networks with node attraction. Moreover, the non-homogenous Poisson model, which is a special case of the doubly stochastic process, can be used to approximate hard-core networks (a model with node repulsion) as described later in this paper.

\subsection{Main Contributions}

In this paper, we develop a framework to analyze the Signal-to-Interference-plus-Noise Ratio (SINR) of a representative link with a multiantenna linear Minimum-Mean-Square-Error (MMSE) receiver in networks where nodes are distributed in space according to a doubly stochastic or Cox process (e.g., see \cite{Stoyan}). This model allows for  non-homogeneity and certain forms of correlation in the spatial node distributions. Two special cases of doubly stochastic processes are non-homogenous PPPs, where node locations are independent but the spatial node distribution is non-uniform, and Poisson cluster processes, where nodes are distributed in clusters whose centers form a PPP. Other examples include networks with a single, randomly located cluster of nodes, and networks with random degrees of clustering. Both non-homogenous PPPs and Poisson cluster networks have been proposed as models for networks with non-homogenous spatial distributions, and analyzed for single antenna systems in works such as \cite{WinNetworkInterference}, \cite{GantiClustered}, \cite{GantiHighSIR}, and
\cite{TanbourgiIsotropic}. Analyzing such network models in systems with multiuser, multiantenna receivers (e.g., the MMSE receiver) is interesting given that almost all existing results consider either  multiuser, multiantenna systems in homogenous PPP networks, or single antenna systems in more general network models. Moreover, multiantenna receivers can suppress interference from nearby nodes and is thus less susceptible to the presence of nearby interferers compared to single antenna systems. 
Compared to non-interference-mitigating receivers (with single or multiple antennas),  the MMSE receiver is generally expected to perform better. This difference is more significant in networks with clusters of nodes than that in spatially homogenous networks.

\begin{figure}[t]
\center
\includegraphics[width = 3in]{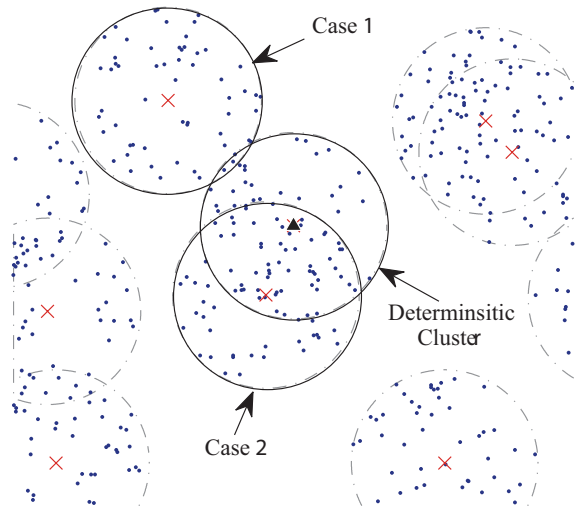}
\caption{Homogenously distributed interferers (points) restricted to a disc at each of the homogenously distributed cluster centers (crosses) with the representative receiver (triangle) at the center of the network. In this example, it is conditioned that there is a cluster at the origin.   Case 1 represents a cluster that does not include the origin, whereas Case 2 represents a cluster that includes the origin.}
\label{Fig:NodesInPlane_MCP}
\end{figure}

We apply the general framework developed in this paper to a number of examples to illustrate its applicability across a wide range of systems. The example network models analyzed here include Poisson cluster networks (an example is shown in Fig. \ref{Fig:NodesInPlane_MCP}) which can model networks with multiple, randomly distributed clusters (proposed as a model for clustered networks in \cite{GantiClustered})  and networks with one, randomly located cluster. Expressions for the CDF of the SINR in these network models are provided in integral form which can be easily evaluated using standard numerical integration techniques. This framework can also be applied to other network models, with further examples provided in conference versions of this paper.

Other network models analyzed here are networks with a single deterministic cluster and an inverse power-law spatial node intensity, for which a closed form expression for the cumulative distribution function (CDF) of the SINR for a receiver at the center of the cluster is given. This result can be used to compare the benefits of using multiple antennas in networks with varying degrees of clustering. We also analyze strip networks where nodes are distributed uniformly randomly on a strip on the plane as a simple model for vehicles on a long roadway.

Furthermore, we apply this model to approximate hard-core point processes which are notoriously difficult to analyze exactly due to the dependence between node locations. In a hard-core process, each node is surrounded by a guard zone which keep nodes at a minimum distance from each other. These processes can be used to model active nodes in  CSMA networks \cite{GantiHighSIR} and  networks where nodes cannot physically come too close to each other. The standard approach to analyze such networks is to approximate them as non-homogenous PPPs with a lower density of nodes around a representative node (e.g., \cite{nguyen2007stochastic, hasan2007guard, HunterMIMOCSMA, ElSawyCSMA}). Existing works however, have not used multiuser receivers such as the MMSE receiver which has the capability to mitigate interference. Moreover these works use a more complicated non-homogenous Poisson approximation based on the second order product density of Matern processes (e.g., see \cite{Stoyan}) compared to the two-case piecewise constant model used here. Since multiuser receivers and hard-cores can both be used to reduce interference, studying MMSE receivers in hard-core networks can help us understand if it is beneficial to use CSMA-like protocols with interference mitigating receivers and vice-versa since both methods can incur significant overhead.

In addition to the different network models, we show that the SINR of a multiantenna receiver  in  wireless networks with more general forms of path-loss than the standard inverse power-law model, is statistically equivalent to the SINR in a network with a non-homogenous Poisson distribution of nodes with an appropriate model for the spatial node distribution. We also show that if the number of receiver antennas is scaled linearly with the nominal node density in doubly stochastic networks, the SINR at a representative receiver converges in distribution to a random variable, and for the case of non-homogenous PPPs, it converges in probability to a positive constant. This finding generalizes similar findings derived previously for homogenous PPPs of nodes in
\cite{govindasamy2007spectral}, \cite{jindal2011multi} and \cite{ali2010performance}. The practical utility of this result is that it indicates that to the extent that system assumptions hold, doubly stochastic networks can be scaled by increasing the number of receiver antennas with the nominal user density. Note that our finding is non-trivial as compared to the results in \cite{govindasamy2007spectral}, \cite{jindal2011multi} and \cite{ali2010performance} which all assumed spatially homogenous and uncorrelated user distributions. In those works, it was possible to find expressions which involve the density of  users and the number of antennas. The double limit as the number of antennas and density of users go to infinity with a constant ratio of these expressions can be taken directly. In this work however, a more complicated approach had to be taken which involved the derivation of an apparently novel property of upper regularized gamma functions given in Lemma \ref{Lemma:RegGammaLimit} below.

In summary the main contributions of this work are
\begin{enumerate}
\item	A general framework to statistically analyze the SINR in doubly stochastic networks including a framework for general construction of doubly stochastic networks. This framework is applicable to a variety of practically relevant scenarios that are interesting problems in their own right.
\item	Proof that scaling number of antennas with nominal user density can help scale doubly stochastic networks.
\item	Exact expressions for the CDFs of the SINR for Thomas and Matern cluster networks with the representative transmitter being part of a cluster of nodes.
\item	Closed-form equations for certain clustered network models to quantify the benefits of multiple antennas in cluster networks.
\item	Simple approximations to the CDF of the SINR for hard-core networks which are difficult to analyze exactly.
\end{enumerate}

\subsection{Related Results}
As remarked earlier, the literature on spatially distributed networks with multiantenna users has focused primarily on networks with nodes distributed as
a homogenous PPP. For such networks, \cite{hunter2008transmission} analyzed the performance of matched-filter and antenna selection receivers, \cite{ali2010performance} analyzed  linear MMSE receivers and \cite{jindal2011multi} analyzed a  partial-zero forcing receiver which includes the standard zero-forcing receiver as a special case. Multiantenna transmitters with spatial multiplexing were analyzed in \cite{govindasamy2007spectral}, \cite{louie2011open}, \cite{GovindasamyTxCSI12} and \cite{vaze2012transmission} under different sets of assumptions. \cite{govindasamy2007spectral}, \cite{louie2011open} and \cite{GovindasamyTxCSI12} also find the optimal number of streams (i.e. multiplexing rate) to maximize the overall data rate in homogenous PPP networks.

Few works have analyzed multiantenna systems in non-homogenous or clustered networks. \cite{TreschCluster} considered interference-alignment in
clustered wireless networks where  partial interference-alignment is used to reduce the system to
a form similar to a single-antenna system.  While interference-alignment can provide enormous data rates, it
requires significant overhead for the exchange of transmit (Tx)
Channel-State Information (CSI). In comparison, the system we
analyze is more attractive for implementation as it does not
require Tx CSI and uses a linear receiver which only requires
CSI of the target transmitter and the spatial covariance matrix
of the interference plus noise. \cite{HunterMIMOCSMA}
approximates networks of multiple-input-multiple-output (MIMO)
links with CSMA using a
Poisson approximation for the spatial node distribution.
However, the multiple antennas are not used for interference
mitigation compared to this work which considers interfering single-input-multiple-output (SIMO) links with interference mitigating
multiantenna receivers.

\cite{NonHomogAsymp}  considers non-homogenous Poisson networks using
an asymptotic analysis which is applicable only with moderately large numbers
of antennas. Furthermore, it does not provide the
distribution of the SINR and focuses on the convergence
of appropriately normalized versions of the SINR as the numbers
of antennas per receiver gets large. In the equivalent
asymptotic regimes,  our results agree with those findings.

In a recent, independent parallel work \cite{ShiRitcey} (which appeared after our conference paper that forms the basis of the results in this work \cite{ZhuNonHomog}) the CDF of the SINR was derived for hierarchical Poisson networks which is used to analyze
Poisson cluster networks.   In \cite{ShiRitcey} the authors assume  that the representative transmitter whose SINR is analyzed,  is located at a deterministic point which is not part of a cluster, even though all other transmitters in the network belong to clusters. For the results on Poisson cluster processes in this paper, the representative transmitter could either be part of a cluster or not, and is thus more general. Moreover, their results differ from ours in that their expressions for the CDFs of the SINR involve complex contour integration whereas corresponding results in our paper involve multiple real integrals.

\subsection{Notation}

Throughout the paper, uppercase bold characters
represent matrices and lowercase bold characters represent
vectors. The indicator function $\mathbf{1}_{\{\mathcal{A}\}}$
equals $1$ if $\mathcal{A}$ is true, and $0$ otherwise. $B(Y, R)$ denotes a disk of radius $R$ centered at $Y$.

\section{System Model}\label{Sec:SystemModel}\label{Sec:ChannelModel}\label{Sec:NodeModel}



A representative receiver at the origin
is communicating with a representative transmitter at a fixed
distance $r_T$ in the presence of simultaneously transmitting co-channel interferers distributed on a plane.
The spatial distribution of interferers will be described later in this section.
These interferers, transmitting with equal
power, are communicating with other receivers whose
locations do not affect our results. We assume the inverse power-law path-loss model where
the average power (over fading realizations) $p$ from a
node transmitting with unit power, at a distance $r$
is $p = r^{-\alpha}$, with the path-loss exponent $\alpha>2$.
The receiver has $L$ antennas, and the representative
transmitter and each interferer have a single antenna. We use
the label $T$ to denote the representative transmitter and
$1, 2, \cdots, n$ to label the interferers.  $r_i$  represents the distance between the
 $i$-th interferer and the representative receiver at the origin, and $x_T$ and $x_i$ represent the transmitted symbols from the representative transmitter and $i$-th interferer respectively. At a given sampling time, the received signal vector $\mathbf{y}\in \mathbb{C}^{L\times 1}$ is
\begin{align}
\mathbf{y} = r_T^{-\alpha/2}\mathbf{g}_T x_T + \sum_{i=1}^n r_i^{-\alpha/2}\mathbf{g}_i x_i + \mathbf{w}\,,
\end{align}
where $r_T^{-\alpha/2}\mathbf{g}_T$ (or
$r_i^{-\alpha/2}\mathbf{g}_i$) represents the channel
coefficients between the representative transmitter (or the
$i$-th interferer) and the receiver. $\mathbf{g}_T$ and
$\mathbf{g}_i \in \mathbb{C}^{L \times 1}$ comprise
independent and identically distributed (i.i.d.), zero-mean,
unit-variance complex Gaussian entries. $\mathbf{w}$ comprises,
i.i.d. complex Gaussian entries with variance $\sigma^2$ per complex dimension, representing noise.

The representative receiver estimates $x_T$ from $\mathbf{y}$
using a linear MMSE estimator which maximizes the
SINR, and which is  given by:
\begin{align}
\text{SINR} = r_T^{-\alpha}\mathbf{g}_T^\dagger \left( \mathbf{G}\mathbf{P}\mathbf{G}^{\dagger}+\sigma^2 \mathbf{I}_L \right)^{-1}\mathbf{g}_T,
\end{align}
where  $\mathbf{I}_L$ is the $L \times L$ identity matrix,
$\mathbf{P} = diag\left[r_1^{-\alpha}, r_2^{-\alpha}, \dotsb
r_n^{-\alpha}\right]$, and the $i$-th column of
$\mathbf{G}\in\mathbb{C}^{L\times n}$ is $\mathbf{g}_i$. To
simplify notation, we define the distance-normalized SINR as
$\gamma = \text{SINR}\cdot r_T^\alpha$.


We assume that the interferers are distributed spatially according to a doubly stochastic process, which is a generalization of the PPP. Doubly stochastic processes can be described by first defining a non-homogenous PPP, which is a point process where node locations are independent, and the number of nodes in any subset $\mathcal{B}$ of the plane is a Poisson random variable with mean
\begin{align}
\mu(\mathcal{B}) = \int_\mathcal{B}\Lambda(r, \theta)\, r\, dr\, d\theta\,.
\end{align}
Here the intensity function $\Lambda(r,\theta)$ captures the likelihood of interferers occurring in an infinitesimal region around a point $(r, \theta)$.  In the doubly stochastic, or Cox process, $\Lambda(r, \theta)$ is a random process (e.g., see \cite{Stoyan}). For a particular realization of the intensity function of the doubly stochastic process, denoted by $\lambda(r,\theta)$,  the process reduces to a non-homogenous PPP. Note here that different models of
spatial node distributions result in different forms for  $\Lambda(r,\theta)$ and $\lambda(r,\theta)$.

Given a deterministic intensity function $\lambda(r,\theta)$, we can construct a non-homogenous PPP of interferers starting with a circular network of radius $R$ and  i.i.d. interferers placed according to the probability density function (PDF) $f_{r,\theta}(r,\theta)$ which is
related to the intensity function as follows:
\begin{align}\label{PDFtoIndFunc}
f_{r,\theta}(r,\theta) = \frac{r}{\mu}\lambda(r,\theta)\mathbf{1}_{\{0 \le r < R\}}\,,
\end{align}
where the number of interferers $n$ in the circular network is a Poisson random variable with mean $\mu$, defined as
\begin{align}   \label{Eqn:Mean}
\mu = \int_0^R\int_0^{2\pi}r\lambda(r,\theta)d\theta dr.
\end{align}
In the derivation of the main results we take $R \to \infty$ to
model the interferers distributed according to a
non-homogenous PPP with intensity function $\lambda(r,
\theta)$.

Note that since the spatial distribution of interferers in our network model is not necessarily homogenous, the representative receiver does not correspond directly to the notion of the ``typical" receiver commonly encountered in the literature (e.g., see \cite{HaenggiJSAC} and references therein) because unlike homogenous networks, statistical properties of the system at any point on the plane (e.g., the origin) could
differ from the properties at other points. For the purposes of this work, the representative receiver should be interpreted
simply as the receiver at the origin, and the representative transmitter is the transmitter to which it is linked.

\section{General Results on the Outage Probability}\label{Sec:MainResults}

One key performance measure of communication systems is the outage  probability, which is defined as the
probability that the SINR is below a threshold $\tau$. For a
fixed $r_T$, this probability is $\text{Pr}\{\text{SINR} \le
\tau\}= F_\gamma(\tau r_T^\alpha)$, where $F_\gamma(\gamma)$ is
the CDF of the distance-normalized SINR $\gamma$.


To characterize the SINR with doubly stochastic processes of interferers, we first condition on a realization of the intensity function, then find the outage probability in the resulting  non-homogenous PPP, and finally  remove the conditioning to derive the outage probability. The following lemma characterizes the SINR when we condition on a realization  $\lambda(r, \theta)$, of $\Lambda(r, \theta)$.  Note that this lemma first appeared as Theorem 1 in \cite{ZhuNonHomog} which is a conference version of this paper, and  a similar result appeared later in another work \cite{ShiRitcey}.

\begin{lemma}\label{Theorem:GenCDF}
The CDF of  $\gamma$  conditioned on $\Lambda(r, \theta) = \lambda(r, \theta)$ (resulting in a non-homogenous PPP with intensity function $\lambda(r, \theta)$)  is
\begin{align} \label{Eqn:GenCDF}
F_{\gamma |\Lambda}(\gamma|\Lambda = \lambda) &=  1-\sum_{k=0}^{L-1}\frac{(\psi(\gamma; \lambda)+\sigma^2\gamma)^k}{k!}\exp(-\psi(\gamma; \lambda)-\sigma^2\gamma)  =  1-\frac{\Gamma(L, \psi(\gamma; \lambda)+\sigma^2 \gamma)}{\Gamma(L)} 
\end{align}
where
\begin{align}\label{Eqn:Psi}
\psi(\gamma; \lambda) =  \int_0^\infty \int_0^{2\pi} \lambda(r,\theta)r\frac{r^{-\alpha}\gamma}{1+r^{-\alpha}\gamma}d\theta dr,
\end{align}
 and $\Gamma(.)$ and $\Gamma(.,.)$ are the gamma function and the upper incomplete gamma function. In addition, the corresponding PDF of  $\gamma$ is:
\begin{align*}
 f_\gamma(\gamma) =  \frac{(\psi(\gamma)+\sigma^2\gamma)^{L-1}\exp(-\psi(\gamma)-\sigma^2\gamma)(\sigma^2+\psi^{\prime} (\gamma))}{(L-1)!}
\end{align*}
where $\psi^{\prime} (\gamma)$ is the  derivative of
$\psi(\gamma)$ with respect to $\gamma$.
\end{lemma}

{\it Proof:} Given in Appendix \ref{Sec:ProofOfGenCDF}.

This result can also be used directly if the random intensity function equals a deterministic function (i.e. $\Lambda(r, \theta) = \lambda(r, \theta)$) with probability 1. We apply this lemma to characterize the outage probability in doubly stochastic networks in the following theorem.

\begin{theorem}\label{Theorem:GenCDFCox}
The CDF of  $\gamma$ in a network with interferers distributed according to a doubly stochastic process is
\begin{align} \label{Eqn:GenCDFCox}
F_{\gamma}(\gamma) = 1 - \sum_{k=0}^{L-1}\frac{\exp\left(-\sigma^2\gamma\right) }{k!} \sum_{\ell = 0}^{k} {k \choose \ell}(\sigma^2\gamma )^{k-\ell} \mathbf{E} _{\Lambda}\left[ \psi^\ell(\gamma; \lambda) \exp(-\psi(\gamma; \lambda))  \right],
\end{align}
where $\psi(\gamma; \lambda)$ follows from \eqref{Eqn:Psi} and $ \mathbf{E}_\Lambda$ denotes taking the expectation over all realizations of $\Lambda(r,\theta)$, i.e. all possible $\lambda(r,\theta)$.
\end{theorem}

{\it Proof:} 
Taking the expectation of \eqref{Eqn:GenCDF} over $\Lambda(r,\theta)$, expanding $(\psi(\gamma;\lambda)+\sigma^2\gamma)^k$ using the binomial theorem, and rearranging the terms yields \eqref{Eqn:GenCDFCox}.

Suppose that the doubly stochastic process of interferers is the superposition of a non-homogenous PPP with intensity function $\lambda_p(r,\theta)$ and another doubly stochastic point process with intensity function $\Lambda_q(r, \theta)$, which results in $\Lambda(r, \theta) = \lambda_p(r,\theta) + \Lambda_q(r,\theta)$.  Substituting this  intensity function into Theorem \ref{Theorem:GenCDFCox} and moving the integral involving $\lambda_p(r,\theta)$ outside of the expectation results in the following corollary.

\begin{corollary}\label{Corollary:GenCDFCox}
\begin{align} \label{Eqn:GenCDFCoxCol}
F_{\gamma}(\gamma) = 1 - \sum_{k=0}^{L-1}\frac{\exp\left(-\psi_p(\gamma)-\sigma^2\gamma\right) }{k!} \sum_{\ell = 0}^{k} {k \choose \ell}(\psi_p(\gamma)+\sigma^2\gamma )^{k-\ell} \mathbf{E} _{\Lambda_q}\left[ \psi_q^\ell(\gamma) \exp(-\psi_q(\gamma))  \right],
\end{align}
where $\psi_p(\gamma)$ and $\psi_q(\gamma)$ are given by \eqref{Eqn:Psi}, with $\lambda_p(r, \theta)$ and $\lambda_q(r, \theta)$ replacing $\lambda(r,\theta)$ respectively. $\lambda_q(r, \theta)$ here denotes a realization of $\Lambda_q(r, \theta)$.
\end{corollary}

This corollary is useful later in this paper in characterizing networks with clusters of nodes conditioned on the location of one or more clusters, which is useful to characterize networks where the representative transmitter at a given point belongs to a cluster.

\section{Scaling Doubly Stochastic Networks by Increasing the Number of Antennas} \label{Sec:Scaling}

One of the questions that Theorem \ref{Theorem:GenCDFCox}
allows us to answer is
whether one can maintain a non-zero Signal-to-Interference-Ratio (SIR) if the number of
antennas at the representative receiver is increased linearly
with the nominal density of interferers in the network. In the
context of homogenous networks \cite{govindasamy2007spectral},
\cite{jindal2011multi} and \cite{ali2010performance} found that
this is indeed the case. Here, we shall show that a similar
result holds even when the spatial interferer distribution is doubly stochastic.

Assuming that noise is negligible, we show that the SIR on the
representative link converges in distribution if the
number of antennas at the receiver increases linearly with a
nominal interferer density. This is under the assumption that
the channel model, independent Rayleigh fading in particular,
holds, and that accurate measurements of CSI are available at
the receiver. A key result that we use is the following lemma
which may already be known but we were unable to find it in the
literature.
\begin{lemma} \label{Lemma:RegGammaLimit}
Let the upper regularized gamma function  be denoted by  $Q(L,x)=\frac{\Gamma(L,x)}{\Gamma(L)}$, where $\Gamma(L, x)$ is the upper incomplete gamma function.  Let $L$ be a positive integer and $q > 0$, then
\begin{align} \label{Eqn:Lemma1}
\lim_{L\to\infty} Q(L, q L) = \begin{cases} 0, & \text{if $q \ge1$}
\\
1, &\text{if $q < 1$}.
\end{cases}
\end{align}
\end{lemma}

{\it Proof:} Given in Appendix \ref{Sec:ProofOfRegGammaLimit}.
Note that the proof here is a corrected version of the proof of
Lemma 1 in a conference version of this paper,
\cite{ZhuNonHomog}.

Suppose that the intensity function $\Lambda(r,\theta) =
\beta\Lambda_c(r,\theta)$, where $\Lambda_c(r,\theta)$ is a
nominal intensity function which describes the ``shape" of the
true intensity function, and  $\beta$ is the nominal interferer
density which scales the nominal intensity function. Let $\lambda(r,\theta)$ be a realization of $\Lambda(r,\theta)$ and $\lambda_c(r,\theta)$ be a realization of $\Lambda_c(r,\theta)$, such that $\lambda(r,\theta) = \beta\lambda_c(r,\theta)$. Next,
define:
\begin{align} \label{Eqn:PsiC}
\psi_c(\gamma; \lambda_c) =   \int_0^\infty \int_0^{2\pi} \frac{\lambda_c(r,\theta)}{\beta}r\frac{r^{-\alpha}\gamma}{1+r^{-\alpha}\gamma}d\theta dr\,.
\end{align}
Note that $\psi(\gamma; \lambda)$ in \eqref{Eqn:Psi} is equal to $\beta \psi_c(\gamma; \lambda_c)$. We can now state the following theorem.
\begin{theorem}\label{Theorem:CDFConverge}
Let $\beta = \ell L$ with a constant scaling coefficient $\ell
> 0$. As $L \rightarrow \infty $, the distance-normalized SIR, $\gamma$ converges in distribution to a random variable with CDF $\mathbf{E}_\Lambda\left[\phi(\gamma;\lambda)\right]$, where
\begin{align}
\phi(\gamma; \lambda) =  \begin{cases} 0,  & \text{if $\gamma \le \psi_c^{-1}\left(\frac{1}{\ell };\lambda\right) $}      \\     1, &\text{if $\gamma > \psi_c^{-1}\left(\frac{1}{\ell };\lambda\right)$}.
\end{cases}
\end{align}
For a $\Lambda(r,\theta)$ that is equal to a deterministic intensity function with probability 1 (i.e. the interferers form a non-homogenous PPP on the plane), the SIR converges in probability to
$\psi_c^{-1}\left(\frac{1}{\ell}\right)r_T^{-\alpha}$.
\end{theorem}

{\it Proof:} Given in Appendix \ref{Sec:ProofOfCDFConverge}.

Therefore, if we increase the number of antennas linearly with
the nominal interferer density in a non-homogenous Poisson
network, the SIR will approach a constant non-zero value. For general doubly stochastic networks, the SIR approaches a random variable with a CDF that is dependent on the statistical properties of the intensity function  which is a random process.  This
fact implies that such networks can be scaled by linearly
increasing the number of antennas per receiver with node
density without degrading the SIR to zero, provided that the
assumptions of the system are  satisfied. Note that as the
number of antennas gets very large, the independent Rayleigh
fading and accurate receiver CSI assumptions will
require increased antenna separations and increased channel
estimation times. An application of this result is presented in Section \ref{Sec:PiecewisePowerLaw}.


\section{Single Cluster Networks} \label{Sec:RandomCluster}
The doubly stochastic network model can be applied to model cluster networks where the cluster centers or the receiver is randomly located. In this section, we consider the scenarios where there is exactly one cluster in the network. Here we shall fix the location of the representative receiver at the origin but with a randomly located cluster. Note that a randomly located receiver with a fixed cluster could also be analyzed using this technique as we are only concerned with the relative locations of the receiver and the cluster.

We assume that the interferers are clustered around a randomly-located parent point (which is not an interferer), $X_0$.  We denote the  PDF of the cluster center by  $f_{X_0}(r,\theta)$.
Conditioned on $X_0$, the daughter points follow a non-homogenous PPP with intensity function $\lambda(r,\theta; X_0)$ which is related to $f_{X_0}(r,\theta)$ by
\begin{align}\label{PDFtoIndFunc}
f_{r,\theta}(r,\theta|X_0) = \frac{r}{\mu_d}\lambda(r,\theta; X_0).
\end{align}

Applying Theorem \ref{Theorem:GenCDFCox}, we find that
\begin{align} \label{Eqn:CDF_Single}
F_\gamma(\gamma) = 1-\sum_{k=0}^{L-1} \int_0^\infty \int_0^{2\pi} \frac{(\psi(\gamma; X_0)+\sigma^2\gamma)^k}{k!} \exp(-\psi(\gamma; X_0)-\sigma^2\gamma) f_{X_0}(\tau,\vartheta) d\tau d\vartheta,
\end{align}
where $\psi(\gamma; X_0)$ is given by $\eqref{Eqn:Psi}$ with $\lambda(r, \theta) = \lambda(r,\theta; X_0)$.

\subsection{Two-dimensional Gaussian Cluster Networks}
%
%
%

Consider a cluster model where the interferers are distributed according to a two dimensional Gaussian distribution (with width parameter $\nu$), centered at $X_0$. This model can be viewed as one with a single cluster from a Thomas Cluster Network analyzed for single antenna systems in \cite{GantiClustered} (referred to as ``a symmetric normal distribution").
Note that under this model, the distance between an interferer to the representative receiver at the origin is Rician distributed. Hence, the PDF of the distance from an interferer to the representative receiver at the origin is:
\begin{align}
f_r(r| X_0) = \frac{r}{\nu^2} \exp\left( \frac{-(r^2+|X_0|^2)}{2\nu^2} \right) I_0 \left(\frac{r |X_0|}{\nu^2} \right), \label{Eqn:OffCenterGaussianPDF}
\end{align}
 $\psi(\gamma;X_0)$  can then be expressed as:
\begin{align}
\psi(\gamma; X_0) = \mu_d  \int_0^\infty f_r(r| X_0) \frac{r^{-\alpha}\gamma}{1+r^{-\alpha}\gamma} dr\,
\end{align}
which yields the CDF of the SINR when substituted  into
\eqref{Eqn:CDF_Single}. Here, $I_0(\cdot)$ is the zeroth-order modified Bessel function of the first kind.

\begin{figure}[htpb]
\center
\includegraphics[width = 3.5in]{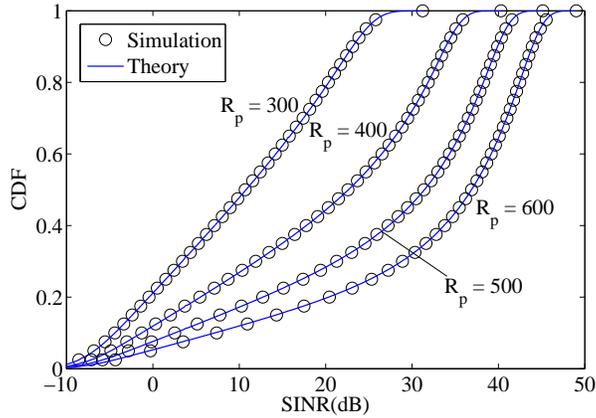}
\caption{Empirical and theoretical CDF of SINR of a  link with 10 receiver antennas, and interferers distributed according to an off-center circular Gaussian. The cluster centers were distributed uniformly in disks of varying radii $R_p$.
\label{Fig:SingleClusterMatch}}
\end{figure}

To verify and illustrate this result (including the accuracy of its numerical evaluation), we simulated this system with  $L = 10$ and $\sigma^2 = 10^{-14}$ .
For each trial, we placed a Poisson number of interferers with mean
$\mu_d = 3140$ in a cluster whose center $X_0$ is distributed with uniform probability in $B(0, X_0)$
The distances of the interferers from the origin thus follow a Rican distribution
with shape parameter $\nu = 100$ for each trial. In Fig.
\ref{Fig:SingleClusterMatch}, the simulated CDFs of the SINR
with $R_p = 300, 400, 500$ and $600$ match our theoretical
predictions  which were numerically evaluated using
standard quadrature integration.

\subsection{Cluster Networks with Power-law Intensity Functions}\label{Sec:PiecewisePowerLaw}
When the location of the cluster center, $X_0$ is a constant with probability 1, the doubly stochastic network reduces to a non-homogenous Poisson network. Non-homogenous Poisson networks with intensity functions in the form of $\rho r^\epsilon$ can effectively model clustered networks where the receiver is located at the center of a deterministic cluster, e.g., \cite{NonHomogAsymp}. The exponent $\epsilon$ determines the degree of clustering and the scale factor $\rho$ is the nominal density. The intensity function includes the homogenous PPP as a special case when $\epsilon = 0$. With a power-law spatial node model, we obtain closed-form expressions for the CDF of the SINR. This finding helps quantify the performance gains that can be expected from using multiple antennas to mitigate interference in the center of a dense cluster as a function of the degree of clustering and other system parameters. To the best of our knowledge, this is the only exact, closed-form result on the CDF of the SINR for a spatially clustered user distribution. 



Here we develop a more general form of such an intensity function. This model will later be used to approximate hard-core networks in Section \ref{Sec:ApproximationofMaternHardCoreProcesswithPiecewiseIntensityFunction}. Consider a set of non-negative numbers representing radial ranges $R_0<R_1<...<R_m$. Assume that the intensity function has the following form:
	\begin{align} \label{Eqn:PiecewiseInt}
	\lambda(r,\theta) = \begin{cases}
 	\rho_1r^{\epsilon_1} & \text{for  $R_0 \le r<R_1$} \\
 	\rho_2r^{\epsilon_2} & \text{for $R_1 \le r<R_2$} \\
             & \vdots\\
	\rho_m r^{\epsilon_m} & \text{for $R_{m-1}\le r<R_m$}\,,\\
	\end{cases}
	\end{align} where $\epsilon_1 > - 2$ if $R_0 = 0$, and
$\epsilon_m<\alpha-2$ if $R_m = \infty$. In the range $R_{k-1}
\le r < R_{k}$, the intensity function of the interferers
follows a power-law distribution with nominal density $\rho_k$,
and exponent $\epsilon_k$. Applying Lemma
\ref{Lemma:HypergeometricFunction}  from Appendix
\ref{Sec:HypergeometricFunction} to \eqref{Eqn:PiecewiseInt},
we find that the CDF of $\gamma$ is given by \eqref{Eqn:GenCDF}
with
	\begin{eqnarray}
	\psi(\gamma) &=& \sum_{k=1}^{m} \frac{2\pi \rho_k }{2+\epsilon_k}\bigg [   R_k^{\epsilon_k+2} {_2F_1}\left( 1, \frac{\epsilon_k+2}{\alpha}; \frac{ \epsilon_k+2+\alpha }{\alpha} ; -\frac{R_k^\alpha}{\gamma} \right)       \nonumber \\ && -  R_{k-1}^{\epsilon_k+2} {_2F_1}\left( 1, \frac{\epsilon_k+2}{\alpha}; \frac{ \epsilon_k+2+\alpha }{\alpha} ; -\frac{R_{k-1}^\alpha}{\gamma} \right)     \bigg ]\,.
	\end{eqnarray}

For the simplest scenario, we have only one piece for the intensity function as follows:
\begin{align}\label{Eqn:PowerLaw}
\lambda(r,\theta) = \rho r^\epsilon,
\end{align}
where $ -2<\epsilon \leq 0$, to maintain finite interference for any $r$. In this case, from the derivation in Appendix F,  the CDF of $\gamma$ is expressible in closed-form (with $\epsilon$ parameterized) as
	\begin{align}\label{Eqn:CDFPowerLaw}
	 F_\gamma(\gamma; \epsilon) =  1&-\sum_{k=0}^{L-1}\frac{1}{k!} \left(\frac{2\pi^2\rho}{\alpha} \csc\left(\pi\frac{\epsilon+2}{\alpha}\right)\gamma^{(\epsilon+2)/\alpha}+\sigma^2\gamma\right)^k\nonumber \\
&\;\;\;\;\;\;\;\;\times \exp \left( - \frac{2\pi^2\rho}{\alpha} 					 \csc\left(\pi\frac{\epsilon+2}{\alpha}\right)\gamma^{(\epsilon+2)/\alpha}-\sigma^2\gamma \right).
	\end{align}
Note that an asymptotic analysis was used to show convergence
of an appropriately normalized version of the SIR for this
model in
\cite{NonHomogAsymp}. To confirm that the result above agrees
with the conclusions in that work, we first neglect the noise
by setting  $\sigma^2 = 0$. Since the SIR grows without bound as $L\to\infty$, we define a normalized version of the SIR,
$\xi = L^{-\alpha/ (2+\epsilon)}r^\alpha_T  \text{SIR} =
L^{-\alpha/ (2+\epsilon)} \gamma $ as is done in
\cite{NonHomogAsymp} to avoid degenerate results as $L\to\infty$. Then
\begin{eqnarray}
 F_\xi(\xi) = \text{Pr}(L^{-\alpha/ (2+\epsilon)} \gamma<\xi) = \text{Pr}(\gamma<\xi  L^{\alpha/ (2+\epsilon)})  = 1-\frac{\Gamma(L, \frac{2\pi^2\rho}{\alpha} \csc\left(\pi\frac{\epsilon+2}{\alpha}\right) \xi^{(\epsilon+2)/\alpha} L)}{\Gamma(L)}\,.
\end{eqnarray}
Given Lemma \ref{Lemma:RegGammaLimit}, if we set
$q =\frac{2\pi^2\rho}{\alpha}
\csc\left(\pi\frac{\epsilon+2}{\alpha}\right)
\xi^{(\epsilon+2)/\alpha}$, then $F_\xi(\xi)$ approaches a step at a deterministic value $  \left[
\frac{2\pi^2\rho}{\alpha}
\csc\left(\pi\frac{\epsilon+2}{\alpha}\right)
\right]^{-\alpha/(\epsilon+2)} $ as $L \rightarrow \infty$. This implies that for large number of antennas $L$,  $\text{SIR} \approx
 \left[ \frac{2\pi^2\rho}{\alpha L}\csc\left(\pi\frac{\epsilon+2}{\alpha}\right)\right]^{-\alpha/(\epsilon+2)} r_T^{-\alpha} $ which is consistent with
the findings in \cite{NonHomogAsymp}.

\begin{figure}
\center
\includegraphics[width = 3.8in]{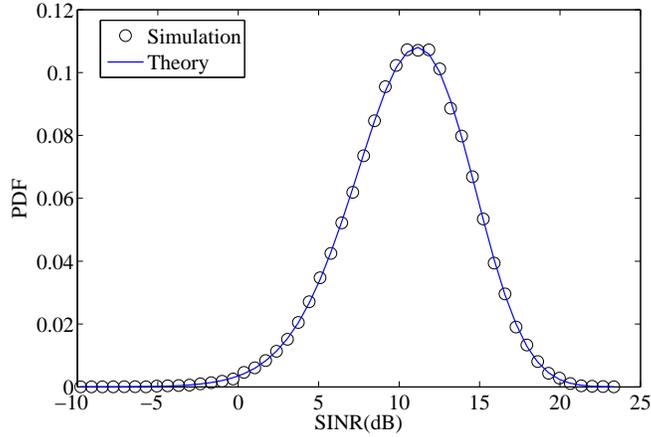}
\caption{Comparison between the empirical and theoretical probability density function of SINR with the power-law intensity function $\Lambda(r,\theta) = \frac{0.023 }{\sqrt{r}}$ . The parameters used are $r_T = 10$, $L = 10$,  $\alpha = 4$, $\sigma^2 = 10^{-12}$, and $100,000$  Monte-Carlo trials.}
\label{Fig:PowerLaw}
\end{figure}

To validate \eqref{Eqn:CDFPowerLaw}, we conducted Monte Carlo simulations which indicate a close agreement between the simulations and the theoretical prediction as illustrated in Fig. \ref{Fig:PowerLaw} which shows  PDFs  of the SINR for the intensity function
$\Lambda(r,\theta) = \frac{0.023 }{\sqrt{r}}$.

It is worth noting that \eqref{Eqn:CDFPowerLaw} can be extended to networks with random degrees of clustering, which are doubly stochastic but not Poisson networks. For instance, consider a  network which is clustered with $\epsilon = \epsilon_1$ with a certain probability, and homogenous otherwise, which could model networks with varying user distributions influenced by usage patterns. The CDF of the SINR in the center of such a network (the location with the worst-case SINR) could be easily found in closed-form from \eqref{Eqn:CDFPowerLaw} and Theorem \ref{Theorem:GenCDFCox} as $\Pr(\epsilon =  \epsilon_1)F_\gamma(\gamma;  \epsilon_1)+(1-\Pr(\epsilon =  \epsilon_1))F_\gamma(\gamma; 0)$.

\begin{figure}[htp]
\center
\includegraphics[width = 4in]{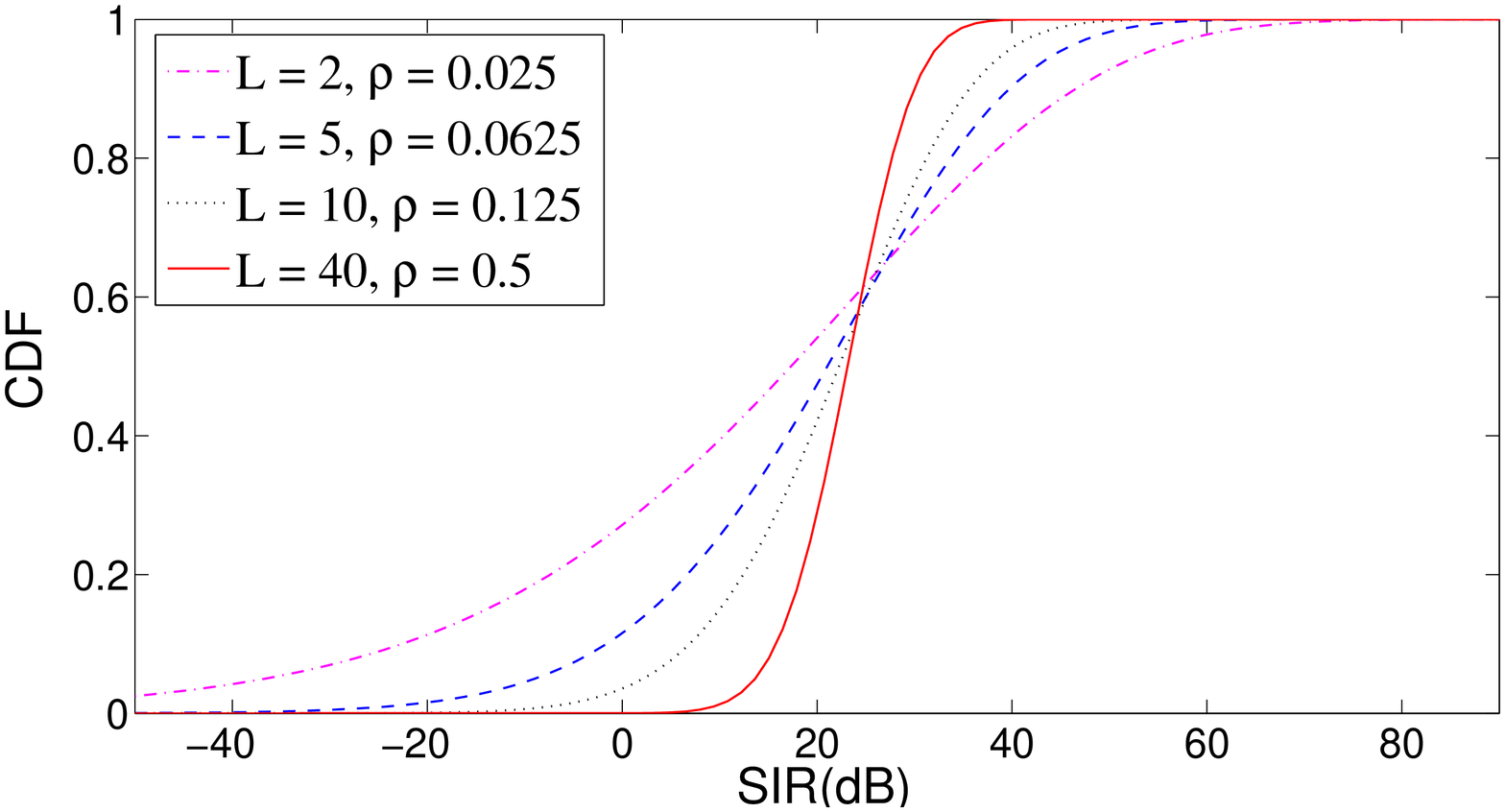}
\caption{Cumulative distribution function of the SIR (dB) with number of antennas increasing linearly with nominal interferer density with $\lambda(r,\theta) = \rho/\sqrt{r}$, $r_T = 10$, $\alpha = 4$ and  $\sigma^2 = 0$.}
\label{Fig:scaled_rho}
\end{figure}

Furthermore, we can use the power-law model as an example to illustrate the scaling properties of the system in Theorem \ref{Theorem:CDFConverge}. We compare the CDF of the SIR with $\lambda(r, \theta) = \beta \lambda_c(r,\theta)$, where $\beta = \rho$ and $\lambda_c(r,\theta) = r^{\epsilon}$ for $L = $2, 5, 10 and 40 antennas, with corresponding density $\rho = $ 0.025, 0.0625, 0.125 and
0.5 in the network. Note that these values correspond to a
linear increase in interferer density with the number of
antennas. The CDFs are illustrated in Fig. \ref{Fig:scaled_rho} which shows that as the number of interferers increases from
$2$ to $20$ with a corresponding increase in the density of interferers, the CDF of SIR approaches a step function, i.e the SIR approaches a
\emph{deterministic} non-zero value in distribution. Moreover, the SIR converging in distribution to a constant implies
that it converges in probability.

\section{Superposition of Poisson and Neyman-Scott Networks}\label{Sec:PoissNSNwks}

In this section, we derive the CDF of the SINR on a representative link
in a network with interferers distributed as a superposition of a non-homogenous PPP and the Neyman-Scott
cluster process. The Neyman-Scott process is  often used in spatial statistics to model random clustering and has been proposed as a model for wireless networks with random clusters of users \cite{GantiClustered}. The superposition of the two processes enables us to analyze a clustered point process of transmitters, conditioned on the location of the cluster containing the representative transmitter. The conditioning on the location of this cluster in turn enables us to condition on the location of the representative transmitter.  An example using the Matern cluster process is given later in this section.

We assume that one subset of interferers is distributed according to a PPP with a deterministic intensity function $\lambda_{p}(r,\theta) $ while the rest are distributed according to the Neyman-Scott cluster process as follows.
A set of parent points, denoted by $\Pi = \{X_i | \ i = 1, 2, ...\}$, is generated from a  PPP with intensity $\lambda^\ast (r,\theta)$ on the plane (see e.g., \cite{Stoyan}), and for each cluster a random number of daughter points are  placed in an i.i.d. fashion according to some probability distribution. To define the Neyman-Scott process within our context, we start with the deterministic intensity function associated with a single cluster whose parent point (or cluster center) is $X_i$, defined as $\lambda(r, \theta; X_i)$.  The number of the i.i.d daughter points that surround parent point $X_i$  is a Poisson random variable with mean $\mu_d$. The relationship between the PDF of the daughter points associated with a parent point  $X_i$  and the intensity function of the parent points $\lambda^\ast (r,\theta)$  depends on the clustering model.  All the daughter points are considered interferers in the network, but the parent points are not. Thus, the intensity function of the interferers in the network, conditioned on a particular realization of the parent point
process $\Pi$, is:
\begin{align} \label{Eqn:LambdaCluster}
\lambda(r, \theta; \Pi) =\lambda_{p}(r,\theta)  +   \sum_{X_i \in \Pi}\lambda(r, \theta; X_i).
\end{align}
\noindent When the conditioning on the realization of the
parent point process is removed we obtain a superposition of Poisson and Neyman-Scott cluster processes.

\subsection{Outage Probability} \label{Sec:CDFCluster}

We can apply Corollary \ref{Corollary:GenCDFCox} to derive the outage probability  for this model. The  expectation  over the random intensity functions $\Lambda(r, \theta)$  can be evaluated using the following Lemma that may already be known, but we were not able to find in the literature.

\begin{lemma} \label{Lemma:Campbell}
Let $\Pi$ be a PPP on $\mathbb{R}^2$ with intensity
$\lambda^\ast: \mathbb{R}^2 \rightarrow [0, \infty)$, and let $\Xi = \sum_{X\in\Pi} \zeta(X)$, where $\zeta:
\mathbb{R}^2 \rightarrow [0, \infty)$ is a non-negative
measurable function. Then, for  integers $\ell\geq 0$,
\begin{align}
\mathbf{E} _{\Pi}\left[\Xi^\ell e^{-\Xi}\right] = & \ell  \exp \left\{ \int_{\mathbb{R}^2} \left(   e^{- \zeta(u)}-1  \right) \lambda^\ast(u) du \right\}   \nonumber \\
&\;\; \times \sum_{(m_1,...,m_\ell) \in \mathcal{M}_\ell} \frac{\prod_{j=1}^{\ell} \left[ \int_{\mathbb{R}^2} \zeta^{j}(z) e^{- \zeta (z)}\lambda^\ast (z)   dz \right]^{m_j}}{m_1! 1!^{m_1}m_2! 2!^{m_2}...m_\ell! \ell!^{m_\ell}},\label{Eqn:Lemma2}
\end{align}
where $u$ and $z$ are integration variables, and $\mathcal{M}_\ell$  is the set of all $\ell$-tuples of nonnegative integers $(m_1,..., m_\ell)$ satisfying the constraint:
\begin{eqnarray}
1\cdot m_1 + 2\cdot m_2 + 3 \cdot m_3 + ...+ \ell \cdot m_\ell = \ell.
\end{eqnarray}
Note that $\zeta^{j}(z)$ refers to the $j$-th power of the
function $\zeta(z)$.
\end{lemma}
{\it Proof:}  Given in Appendix \ref{Sec:Campbell}.

To express the CDF of the SINR on the representative link, we first define
\begin{eqnarray} \label{Eqn:Zeta}
\zeta(X_i) = \int_{0}^{\infty} \int_{0}^{2\pi} \lambda(r, \theta; X_i) r \frac{r^{-\alpha}\gamma}{1+r^{-\alpha}\gamma} d\theta dr,
\end{eqnarray}
and conditioned on the parent point process $\Pi$,  $\psi_q(\gamma)$ in Corollary \ref{Corollary:GenCDFCox} can be expressed as:
\begin{eqnarray} \label{Eqn:PsiPi}
\psi_q(\gamma) =\int_{0}^{\infty} \int_{0}^{2\pi}\sum_{X_i \in \Pi} \lambda(r, \theta; X_i) r \frac{r^{-\alpha}\gamma}{1+r^{-\alpha}\gamma} d\theta dr  = \sum_{X_i \in \Pi}  \zeta(X_i).
\end{eqnarray}
The interchange of the order of integration and summation
follows from Theorem 11.30 in \cite{Rudin}. With this definition, we can state the following result.

\begin{theorem}\label{Theorem:CDFCluster}
The CDF of the distance-normalized SINR $\gamma$ in a network with interferers
distributed as a superposition of a non-homogenous PPP and a Neyman-Scott cluster process is:
\begin{align}
&F_\gamma(\gamma) = 1-\exp \left\{\int_0^{2\pi} \int_0^\infty  \left( e^{- \zeta (r)}-1  \right) \lambda^\ast(r,\theta) r drd\theta \right\} \sum_{k=0}^{L-1} \exp\left(-\psi_p(\gamma)-\sigma^2\gamma\right)   \nonumber \\
&\;\;\;\;\;\;\times \sum_{\ell = 0}^{k} \frac{(\psi_p(\gamma)+\sigma^2\gamma )^{k-\ell}}{(k-\ell)!}   \sum_{(m_1,...,m_\ell) \in \mathcal{M}_\ell} \frac{ \prod_{j=1}^{\ell}    \left[\int_0^{2\pi} \int_0^\infty \zeta^{j}(r) e^{- \zeta (r)}  \lambda^\ast(r,\theta) r  dr d\theta \right]^{m_j} }{m_1! 1!^{m_1}m_2! 2!^{m_2}...m_\ell! \ell!^{m_\ell}}. \label{Eqn:CDFCluster}
\end{align}
\end{theorem}

{\it Proof:}  Evaluating $\mathbf{E} _{\Pi}\left[ \psi_q^\ell(\gamma) \exp(-\psi_q(\gamma))\right]$  in Corollary \ref{Corollary:GenCDFCox} where $\psi_q(\gamma)$ follows from \eqref{Eqn:PsiPi} with Lemma \ref{Lemma:Campbell}, algebraic manipulations, and converting the integrals from Cartesian to polar coordinates yields  \eqref{Eqn:CDFCluster}.

In \eqref{Eqn:CDFCluster}, $\psi_p(\gamma)$ captures the effect of the deterministic portion of the intensity function, $\lambda^\ast(r,\theta)$
represents the intensity function of the parent point process, and $\zeta (r)$ captures the effect of the
daughter points associated with a parent point that is at a distance $r$ from the origin.
For representative Neyman-Scott processes, such as the Matern cluster process, the integrals in equation \eqref{Eqn:CDFCluster} can be evaluated numerically  using standard methods.

\subsection{Application to the Matern Cluster Process Conditioned on a Deterministic Cluster} \label{Sec:MaternCluster}

Here, we apply  Theorem \ref{Theorem:CDFCluster} to analyze the Matern cluster process (a type of Neyman-Scott process), conditioned on a deterministic cluster centered at the origin. Neyman-Scott processes have been used as models for wireless networks with clustered interferers in works such as
\cite{GantiClustered}. Consider Fig. \ref{Fig:NodesInPlane_MCP} which illustrates a realization the superposition of the Matern cluster process and a deterministic PPP where each cluster is a disk of radius $R_d$. The corresponding per-cluster intensity function is
\begin{align}\label{Eqn:MaternSingleClusterIntensity}
\lambda(r,\theta; X_i) = \rho_d \mathbf{1}_{\{(r,\theta)\in B(X_i, R_d)\}}\,,
\end{align}
where $\rho_d = \mu_d/ (\pi R_d^2)$ is the density of the
daughter points in a single cluster. Conditioned on a
 realization of the parent points $\Pi$, the intensity function becomes:
\begin{align}
\lambda(r,\theta; \Pi) =  \rho_d' \mathbf{1}_{\{(r,\theta)\in B(0, R_d)\}}  + \sum_{X_i \in \Pi} \rho_d \mathbf{1}_{\{(r,\theta)\in B(X_i, R_d)\}}.
\end{align}
where  $\rho_d' = (\mu_d-1) /(\pi  R_d^2)$ as we reduce the mean number of interferers by 1 at the deterministic cluster to account for the representative transmitter. The first term represents the deterministic cluster and the second term represents the Matern cluster
process conditioned on the realization of the parent points $\Pi$.

From results in \ref{Sec:PiecewisePowerLaw} and Lemma \ref{Lemma:HypergeometricFunction}, the contribution of the deterministic cluster is:
\begin{align}\label{Eqn:DetClust}
\psi_p(\gamma) = \pi \rho_d' R_d^2{_2F_1} \left(1,\frac{2}{\alpha};\frac{2+\alpha}{\alpha};-\frac{R_b^\alpha}{\gamma}\right) = (\mu_d-1) {_2F_1} \left(1,\frac{2}{\alpha};\frac{2+\alpha}{\alpha};-\frac{R_b^\alpha}{\gamma}\right).
\end{align}

Next, we find the parameters and functions that capture the interferers from the Matern cluster process. To find the corresponding PDF, conditioned on one parent point $X_i$, of the distance of a random point from this
cluster to the representative receiver $f_r(r|X_i)$ , we need
to consider two disjoint and independent cases. The cases
correspond to whether the cluster under consideration includes
the origin or not, as shown in  Fig. \ref{Fig:NodesInPlane_MCP}.

Case 1:  $|X_i| \ge R_d$, i.e. when the representative receiver is outside the disk $B(X_i,
R_d)$. In this case, $r$ is the distance between a random point inside a circle of radius $R_d$ and a fixed point outside this circle (at a distance $|X_i|$). This PDF is given in \cite{Mathai}
as follows:
\begin{align}\label{Eqn:MaternDistanceDistributionCase1}
 f_r(r| \ |X_i| \ge R_d) =   \frac{2  r}{\pi R_d^2} \cos^{-1}\left( \frac{r^2+|X_i|^2-R_d^2}{2 r|X_i| } \right) \cdot \mathbf{1}_{\{ |X_i|-R_d  \le r\le |X_i|+R_d \}}
\end{align}

Case 2: $|X_i| < R_d$, i.e. when the representative receiver is inside disk $B(X_i, R_d)$. 
Using  geometric arguments and applying the techniques used to derive \eqref{Eqn:MaternDistanceDistributionCase1} in  \cite{Mathai} , we found the PDF of $r$ in a similar form as follows:
\begin{align}\label{Eqn:MaternDistanceDistributionCase2}
 f_r(r| \ |X_i| < R_d) =  \frac{2r}{\pi R_d^2}\cdot \mathbf{1}_{\{ r\le R_d - |X_i|\}}
+ \frac{2  r}{\pi R_d^2} \cos^{-1}\left( \frac{r^2+|X_i|^2-R_d^2}{2 r|X_i| } \right) \cdot \mathbf{1}_{\{ R_d-|X_i|  \le r\le R_d + |X_i|\}}.
\end{align}

Substituting \eqref{Eqn:MaternDistanceDistributionCase1} and
\eqref{Eqn:MaternDistanceDistributionCase2} into
\eqref{Eqn:Zeta} yields:
\begin{align*}
\zeta(X_i) =  \begin{cases} \rho_d  \int_{|X_i|-R_d}^{|X_i|+R_d}  2r \cos^{-1}\left( \frac{r^2+|X_i|^2-R_d^2}{2 r|X_i| } \right)\frac{r^{-\alpha}\gamma}{1+r^{-\alpha}\gamma}dr & ,  |X_i| \ge R_d\\
\rho_d \left[  \int_0^{R_d-|X_i|}  2r \frac{r^{-\alpha}\gamma}{1+r^{-\alpha}\gamma}dr  +   \int_{R_d-|X_i|}^{R_d+|X_i|}  2r \cos^{-1}\left( \frac{r^2+|X_i|^2-R_d^2}{2 r|X_i| } \right)\frac{r^{-\alpha}\gamma}{1+r^{-\alpha}\gamma}dr \right] & ,  |X_i| <R_d.  \end{cases}
\end{align*}
The CDF is found by substituting this expression and \eqref{Eqn:DetClust} into \eqref{Eqn:CDFCluster}.

We verified the resulting CDF by Monte Carlo simulations.  In each trial of the simulations we start with a homogenous PPP with density $\rho_p$ to model the cluster centers, and place one deterministic cluster centered at the origin. Next we generate a Poisson number of daughter points at the deterministic cluster with density $\rho_d'$  in a disk of radius $R_d$. Then, we independently generate a Poisson number of daughter points with density $\rho_d$, i.i.d. in disks of radius $R_d$, centered at each of other cluster centers.  Even though we could not solve the integrals in the resulting CDFs of the SINR  in closed form, we were able to apply standard quadrature numerical integration methods  to compare the theoretical predictions to the simulated CDFs of the SINR.

\begin{figure}[htbp]
\center
\includegraphics[width = 3.8in]{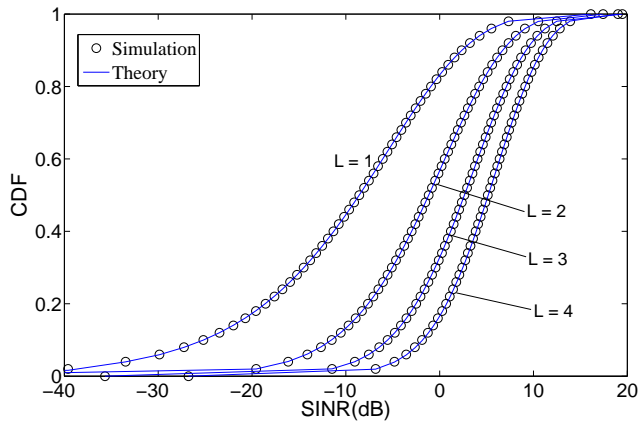}
\caption{Comparison between the empirical and theoretical CDFs of SINR when interferers are generated from a Matern cluster process conditioned on one deterministic cluster. Varying the number of antennas $L$, the empirical CDFs result from $100,000$ Monte Carlo trials. Other parameters used are $r_T = 10$, $\alpha = 4$, $\sigma^2 = 10^{-5}$,  $\rho_p = 1.6 \times 10^{-5}$, $\mu_d = 200$ and $R_d = 300$.}
\label{Fig:MCP_match}
\end{figure}

Fig. \ref{Fig:MCP_match} depicts the simulated and theoretical CDFs of the SINR with interferers distributed as a Matern cluster process, conditioned on a cluster at the origin for $L = 1, 2, 3$ and $4$ antennas, and SNR of 10 dB, and the remaining parameters as specified in the figure.  The simulated and theoretical CDFs of the SINR are indistinguishable, which confirms  the accuracy of both the analysis and the numerical integration. Notice
that the variance of the SINR for the cluster processes is large due to the high degree of irregularity in the spatial
interferer distribution.

\section{Non-homogenous Poisson Networks }\label{Sec:ApplicationstoRepresentativeIntensityFunctions}

As described in Section \ref{Sec:MainResults}, conditioned on a realization of the intensity function, the doubly stochastic process reduces to a non-homogenous PPP, and is as such a special case of a doubly stochastic process. As one of the main results of this section, we show that the SINR of a representative link in a homogenous network with a generalized path-loss model is statistically equivalent to  that in a non-homogenous network with the inverse power-law path-loss model, which enables us to use the framework developed here to characterize systems with more general path-loss models and to compare networks with different path-loss models.

Non-homogenous Poisson networks are interesting as they describe many practical scenarios such as
simple models of roadway networks where the spatial node distribution has a constant positive intensity on the roadway and is zero outside the roadway.
The non-homogenous network model can also be used to approximate the SINR distribution for links in hard-core networks, which serve as simple models for networks with protocols such as CSMA. As noted in the introduction, CSMA networks are notoriously difficult to analyze. In Section \ref{Sec:ApproximationofMaternHardCoreProcesswithPiecewiseIntensityFunction}, we show that a simple approximation of a hard-core process using a non-homogenous PPP can provide an accurate approximation for the CDF of the SINR. Note that existing works on hard core networks, none of which consider interference-mitigating multiantenna receivers as we do here, use more complicated approximations because they are more sensitive to nearby interferers which would be nulled out by the MMSE receiver in our system.




\subsection{Modeling Generalized Path-loss Models through Non-homogenous Poisson Networks}\label{Sec:ArbPathLoss}

In this subsection, we show that the distance-normalized SINR of the representative link in a homogenous PPP network where the path-loss is any monotonically decreasing function of distance, is statistically equivalent to the SINR in non-homogenous PPP network with an appropriate intensity function and the inverse-power-law path loss model. This enables us to apply the framework we developed for the inverse-power law model to analyze networks with a more general form of path loss. Although the inverse power-law model has been experimentally validated for certain physical scenarios (e.g., see \cite{herring2010path}), there are many scenarios for which this model is inadequate as described in \cite{herring2010path} for instance.

Assume that the path-loss is represented by $\varphi(r)$, which is a continuous, monotonically decreasing function.  We note here that the this form of path-loss is quite general and is equivalent to general path-loss models used in \cite{HaenggiJSAC} and numerous other works on spatially distributed networks. The CDF of the distance-normalized SINR $\gamma$ is given in Lemma \ref{Theorem:GenCDF}, and $\Lambda(r,\theta)$ equals some deterministic intensity function $\lambda_G(r,\theta)$ with probability 1. Let $\lambda_G(r,\theta)$ be isotropic in $\theta$ so that it is not dependent on $\theta$. Thus,
\begin{align}
\psi(\gamma) =  2\pi\int_0^\infty  \lambda_G(r) r \frac{\varphi(r)\gamma}{1+\varphi(r)\gamma} dr.
\end{align}

Let $u = [\varphi(r)]^{-\frac{1}{\alpha}}$. Since $\varphi(r)$ is continuous and monotonically decreasing, it is invertible and $r = \varphi^{-1}(u^{-\alpha})$. By change of variables, we have:
\begin{align*}
\psi(\gamma) =  2\pi\int_{[ \varphi(0)]^{-1/\alpha}}^{ [\varphi(\infty)]^{-1/\alpha}}  \lambda_G( \varphi^{-1}(u^{-\alpha})) \cdot  \varphi^{-1}  (u^{-\alpha})\cdot \frac{u^{-\alpha}\gamma}{1+u^{-\alpha}\gamma}\cdot ( \varphi^{-1})'(u^{-\alpha})\cdot(-\alpha)u^{-\alpha-1} du
\end{align*}

We denote a new isotropic intensity function as $\lambda_S(r)$ with
\begin{align}
\lambda_S(r) =  \lambda_G(\ell^{-1}(r^{-\alpha})) \cdot  \ell^{-1}  (r^{-\alpha})\cdot (\ell^{-1})'(r^{-\alpha})\cdot (-\alpha) r^{-\alpha-2}\cdot \mathbf{1}_{ \{ [ \varphi(0)]^{-1/\alpha} \le r \le [ \varphi(\infty)]^{-1/\alpha} \} }.
\end{align}

Hence, we have found an isotropic intensity function $\lambda_S(r)$, under the path-loss modeled by  $ r^{-\alpha}$ such that the resulting CDF of $\gamma$ is \eqref{Eqn:GenCDF}, where
\begin{align}
\psi(\gamma) =  2\pi \int_0^\infty  \lambda_S(r) r \frac{r^{-\alpha}\gamma}{1+r^{-\alpha}\gamma} dr.
\end{align}

So given a certain spatial distribution of interferers under an arbitrary path-loss model $\varphi(r)$, we can find a corresponding spatial node distribution under the path-loss model $r^{-\alpha}$ in our system model that will have the same distance-normalized SINR statistically.

For instance, suppose we have a homogenous network with uniform density $\rho$ and dispersive path loss  $e^{- \nu r}$. This model has been proposed for
certain propagation environments over large distances with supporting data in \cite{Franceschetti2002Microcellular}. Substituting $u = e^{\frac{\nu  r}{\alpha}}$, we get $r = \alpha\ln(u)/ \nu $ and $dr= \alpha/( \nu  u) du$. Therefore, $\psi(\gamma)$  can also be expressed as:
\begin{align}
\psi(\gamma) =   2\pi\int_{\ell^{-\frac{1}{\alpha}}(0)}^{\ell^{-\frac{1}{\alpha}}(\infty)}  \rho \frac{\alpha \ln(u)}{\nu}  \frac{u^{-\alpha}\gamma}{1+u^{-\alpha}\gamma} \frac{\alpha}{\nu u} du  = 2\pi \int_{1}^{\infty} \rho \frac{\alpha^2 \ln(u) }{\nu^2 u^2} u \frac{u^{-\alpha}\gamma}{1+u^{-\alpha}\gamma}du
\end{align}

The equivalent intensity function with path loss $r^{-\alpha}$ is then
\begin{align}
\lambda_S(r) =   \rho \frac{\alpha^2 \ln(r) }{\nu^2 r^2}   \mathbf{1}_{\{ 1 \le r\le \infty \} }
\end{align}

\begin{figure}[tp]
\center
\includegraphics[width = 6.5in]{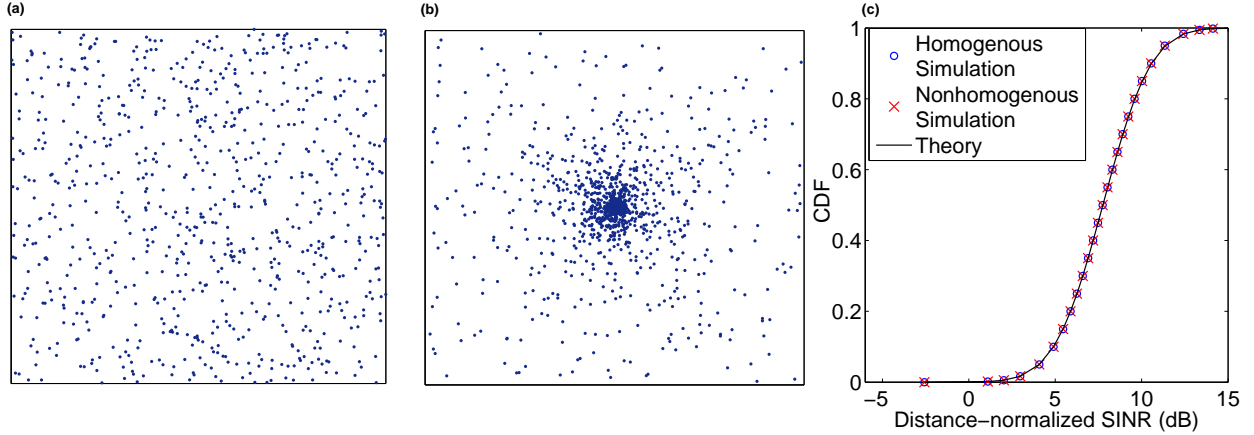}
\caption{ Figure (a) shows a homogenous PPP which with a dispersive path loss model ($e^{- 0.01 r}$), is statistically equivalent at the origin to the non-homogenous PPP in Figure (b) with path loss $r^{-4}$. Figure (c) illustrates simulated and theoretical predictions for the CDF of the SINR for both models, indicating their equivalence. }
\label{Fig:twonetworks}
\end{figure}

We conducted simulations of this model with $L = 10$, $r_T = 10$, $\sigma^2 = 10^{-12}$, a homogenous PPP distribution of interferers and
path-loss: $e^{-\nu r}$, with $\nu = 0.01$, and $\lambda_G(r) = \rho = 10^{-5}$. This model is found to be equivalent to a non-homogenous PPP with  path-loss $r^{-4}$ where, $\lambda_S(r) =   \rho \frac{\alpha^2 \ln(r) }{\nu^2 r^2}   \mathbf{1}_{\{ 1 \le r\le \infty \} } = 1.6 \frac{\ln(r) }{\nu^2 r^2} \mathbf{1}_{\{ 1 \le r\le \infty \} } $. Subplot (c) of Fig. \ref{Fig:twonetworks} shows simulations of networks with both path-loss models along with the theoretical CDF, indicating a close agreement, thereby illustrating that the distance-normalized  SINR for a receiver at the origin in a network with a dispersive path-loss model and the homogenous spatial distribution of nodes shown in subplot (a) of Fig. \ref{Fig:twonetworks}, is equivalent statistically the non-homogenous PPP shown in subplot (b) of the figure, with the $r^{-\alpha}$ path-loss model.

\subsection{Two-dimensional Strip Networks}\label{Sec:NumRoadWay}

\begin{figure}[htp]
\center
\includegraphics[width = 4in]{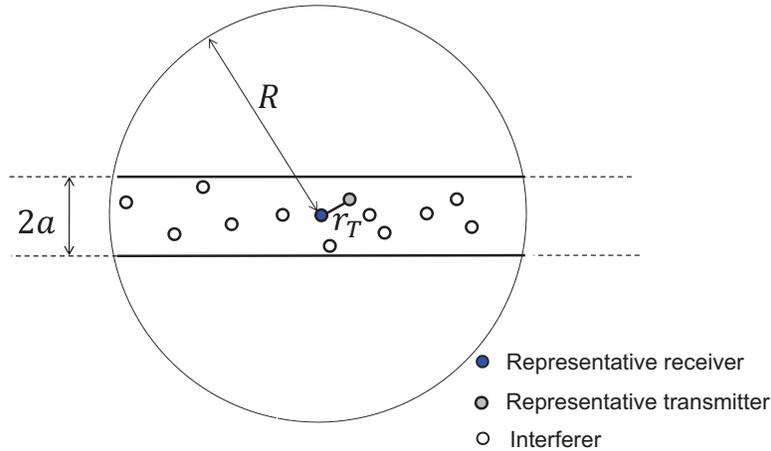}
\caption{Modeling a strip network as a non-homogenous PPP.}
\label{Fig:RoadwayFig}
\end{figure}

We can use the framework for analyzing the SINR in non-homogenous PPPs to characterize a representative link in a strip network, which can be used as a simple model for a vehicular network on a straight roadway \cite{liu2007capacity}. In this model, nodes are distributed according to a non-homogenous PPP on the plane with intensity function equal to $\rho$ in a strip of width $2a$ centered at the origin, and zero outside the strip. Fig. \ref{Fig:RoadwayFig} illustrates an example of such a network. By taking $R\to\infty$, we arrive at an infinite strip network.

Using geometric arguments, we find that the PDF of the distance between the origin and a node distributed with uniform probability in the band-aid-shaped strip illustrated in Fig. \ref{Fig:RoadwayFig} is
\begin{align*}
f_r(r) =
\begin{cases}
2 \pi r/A \  &, \ r \le a \\
4 r \arcsin(a/r)/ A  \ &, \  a < r \le R
\end{cases}
\end{align*}
where $A = 2R^2 \arcsin(a/R) + 2a \sqrt{R^2-a^2}$ is the area of the band-aid shaped strip. As $R\to \infty$, we can write the following
expression in terms of the intensity function $\lambda_b(r, \theta)$:
\begin{align}
\int_0^{2\pi}  r \lambda_b(r,\theta) d\theta = \begin{cases}
 2\rho \pi r \  &, \ r \le a \\
 4 r\rho \arcsin(a/r)  \ &, \  r > a
\end{cases}\,,
\end{align}
We can thus apply  Lemma \ref{Theorem:GenCDF} where $\psi(\gamma)$ in \eqref{Eqn:Psi} becomes:

\begin{align}
\psi(\gamma) &= 2\pi \rho \int_{0}^{a} r  \frac{\gamma}{\gamma+r^{\alpha}} dr + 4\rho  \int_{a}^{\infty}  r\arcsin(a/r)  \frac{\gamma}{\gamma+r^{\alpha}} dr\nonumber \\
&  =\pi \rho a^2 \ \!_2F_1\left(1,\frac{2}{\alpha};\frac{2+\alpha}{\alpha}; -\frac{a^\alpha}{\gamma}\right)  + 4\rho  \int_{a}^{\infty}  r\arcsin(a/r)  \frac{\gamma}{\gamma+r^{\alpha}} dr \label{Eqn:RoadwayArcsin}
\end{align}

\begin{figure}[htp]
\center
\includegraphics[width = 4in]{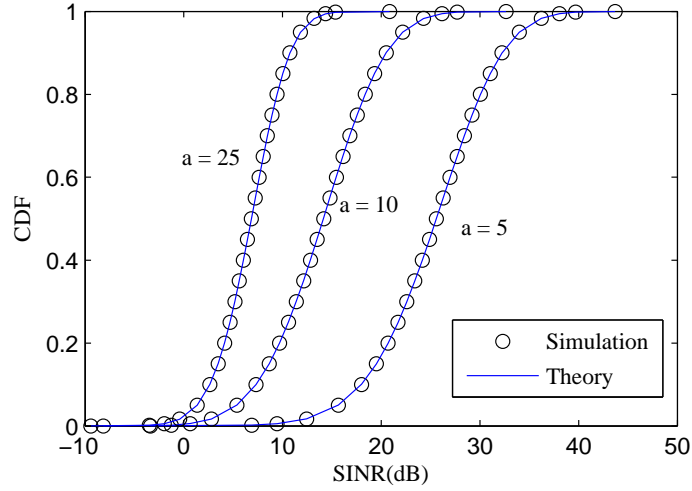}
\caption{Simulated ($10,000$ trials) and theoretical CDF of the SINR (dB) in a road way network with $r_T = 10$, $\sigma^2 = 10^{-12}$, $\alpha = 4$,  $L = 10$, $\rho = 0.01$, and $a =$ 5, 10 and 25 corresponding to the the widths: 10, 20 and 50.}
\label{Fig:RoadWay}
\end{figure}

Fig. \ref{Fig:RoadWay} illustrates results from $10,000$ trials of a Monte Carlo simulation of this network model, with the parameters indicated in the caption. The CDF was evaluated using standard quadrature integration. The close agreement between the simulation and numerical computation validates the results and indicates that accurate numerical evaluation of the CDF is possible using standard techniques.

\subsection{Approximating  Hard-Core Processes by Non-Homogenous Poisson Processes} \label{Sec:ApproximationofMaternHardCoreProcesswithPiecewiseIntensityFunction}

In this section, we use a Poisson approximation to analyze hard core networks.
We construct circular guard zones of radius $R_1$ around all transmitting nodes (representative transmitter and intreferers) which serves as a simplified model for CSMA networks \cite{GantiHighSIR} . The CDF of the SINR for the MMSE receiver with this model can be approximated using a non-homogenous PPP with intensity function
\begin{align}\label{Eqn:OffCenterIntens}
\Lambda_T(r,\theta) = \begin{cases}
 0 & \text{for  $(r,\theta) \in B(X_T, R_1)$} \\
 \rho(R_1). & \text{otherwise} \\
\end{cases}
\end{align}
where $X_T = (r_T, 0)$ is a nominal location of the representative transmitter. The density of interferers is zero in the guard zone of the representative transmitter and equals a constant $\rho(R_1) = \frac{1-\exp(-\rho_p \pi R_1^2)}{\pi R_1^2}$, which is the effective density of nodes from the Matern type-II process, outside the guard zone \cite{Stoyan}. Note that since the Matern hard-core process is isotropic, the angular coordinate of the representative transmitter can be any value here. Using geometric arguments, we found the PDF of the distance from a point outside $B(X_T, R_1)$ to the origin to be
\begin{align}
 f_r(r) =
    \begin{cases}
 	\frac{2\pi r-2 r \cos^{-1}\left(\frac{r^2+r_T^2-R_1^2}{2r r_T}\right)}{A}, & \text{if  $r_T-R_1< r <r_T+R_1$} \\
    \frac{2\pi r}{A}, & \text{otherwise.} \\
    \end{cases}
\end{align}
if $r_T \ge R_1$ (i.e. the representative receiver is outside the guard zone of the representative transmitter), and
\begin{align}
 f_r(r) =
    \begin{cases}
    0, &  \text{if $r < R_1-r_T$ } \\
 	\frac{2\pi r-2 r \cos^{-1}\left(\frac{r^2+r_T^2-R_1^2}{2r r_T}\right)}{A}, & \text{if  $R_1-r_T< r <r_T+R_1$} \\
    \frac{2\pi r}{A}, & \text{otherwise.} \\
    \end{cases}
\end{align}
if $r_T < R_1$ (i.e. the representative receiver is inside the guard zone of the representative transmitter).

Therefore, the CDF is given by \eqref{Eqn:GenCDF} with function $\psi(\cdot)$  evaluated as:
\begin{align}
\psi(\gamma) =&  2\pi \rho(R_1)  \int_{0}^{r_T-R_1} r\frac{r^{-\alpha}\gamma}{1+r^{-\alpha}\gamma}dr  + 2\pi \rho(R_1)  \int_{r_T+R_d}^{\infty}r\frac{r^{-\alpha}\gamma}{1+r^{-\alpha}\gamma}dr \nonumber \\
& +\rho(R_1)  \int_{r_T-R_1}^{r_T+R_d}   \left[ 2\pi r-2 r \cos^{-1}\left(\frac{r^2+r_T^2-R_1^2}{2r r_T}\right) \right] \frac{r^{-\alpha}\gamma}{1+r^{-\alpha}\gamma}dr,
\end{align}
if $r_T \ge R_1$, and
\begin{align}
\psi(\gamma) =& \rho(R_1)  \int_{R_1-r_T}^{r_T+R_d}   \left[ 2\pi r-2 r \cos^{-1}\left(\frac{r^2+r_T^2-R_1^2}{2r r_T}\right) \right] \frac{r^{-\alpha}\gamma}{1+r^{-\alpha}\gamma}dr  \nonumber \\ & +  2\pi \rho(R_1)  \int_{r_T+R_d}^{\infty}r\frac{r^{-\alpha}\gamma}{1+r^{-\alpha}\gamma}dr.
\end{align}
if $r_T < R_1$.

\begin{figure}[htbp]
\centering
\includegraphics[width = 3.5in]{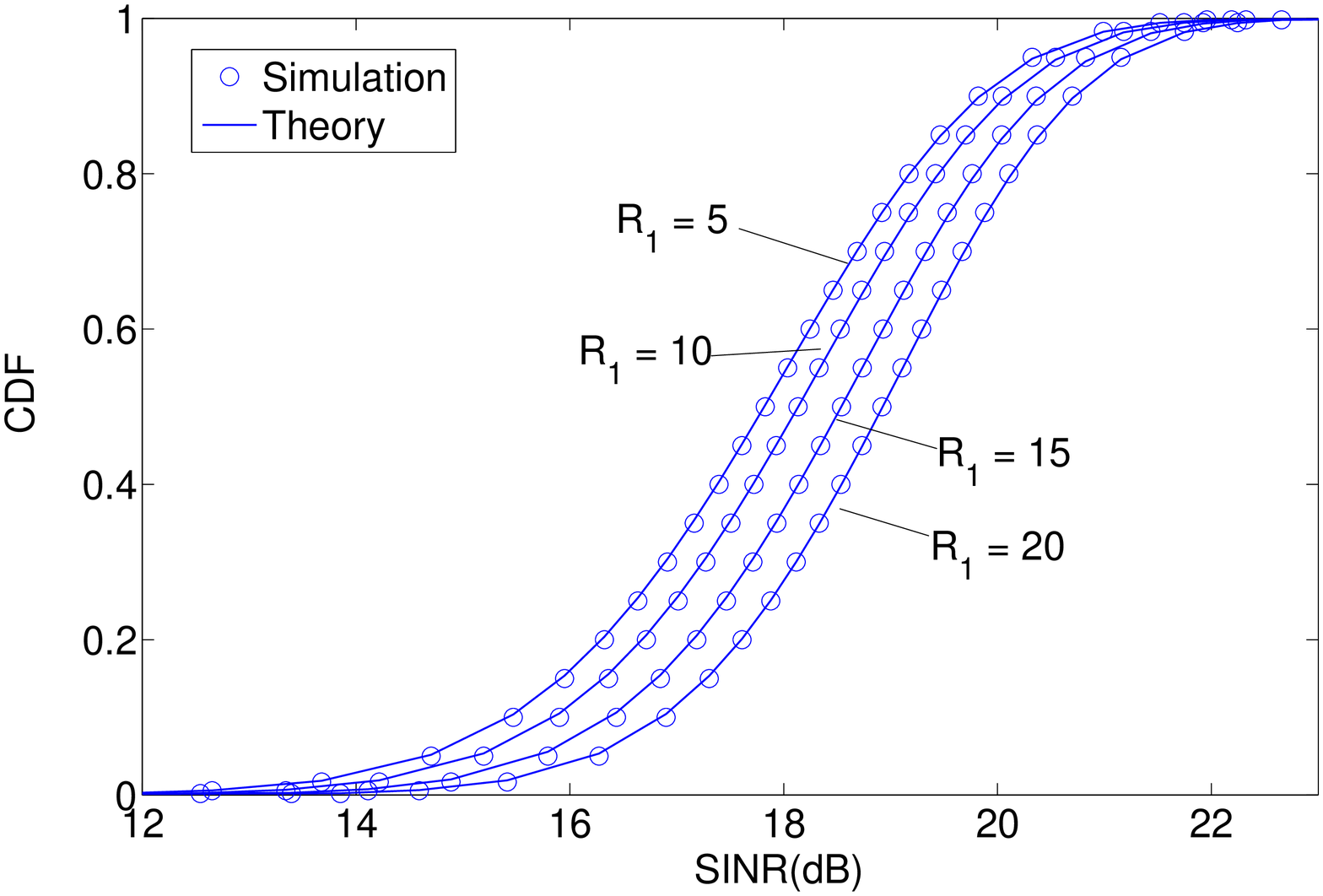}
\caption{CDF of the SINR for Matern type-II networks with hard-cores around all transmitters.}
\label{Fig:OffsenterSINR}
\end{figure}

To verify that this approximation  holds for Matern type-II networks with guard zones around each transmitter, we simulated networks with $L =10, r_T = 10, \alpha = 4$ and SNR = 10 dB  for several different guard zone radii. The results of the simulations and the CDF of the SINR with
$\Lambda(r, \theta)$ from \eqref{Eqn:OffCenterIntens} are show in Fig. \ref{Fig:OffsenterSINR}, which indicates a close agreement between the approximation and the simulated CDF.

Note that in more sophisticated protocols such as CSMA with collision avoidance (CSMA/CA), guard zones are placed around active receivers.
We have found that by centering the guard-zone around the representative receiver instead of the representative transmitter, we can model such networks with comparable accuracy.

\begin{figure}
\centering
\includegraphics[width = 4.2in]{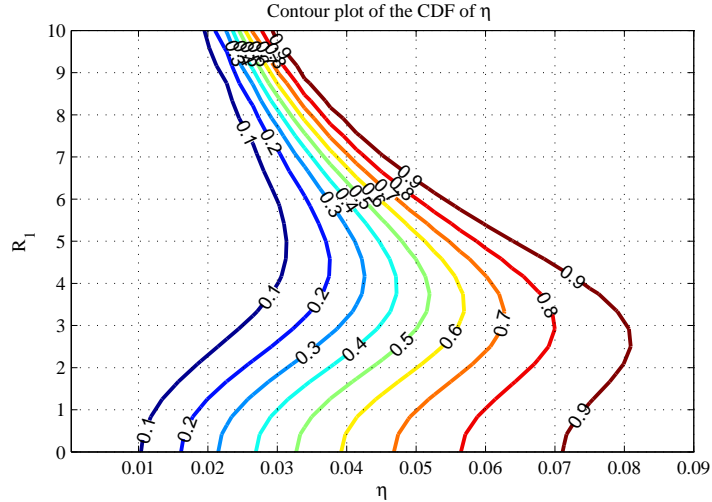}
\caption{Contour plot of the CDF of the spectral efficiency density with respect to the spectral efficiency density $\eta$ and the radius of the guard zone $R_1$. The unit density is $\rho_p = 0.05$ and the distance between the representative receiver and transmitter is $r_T = 5$. .The parameters used are  $\alpha = 4$, $\sigma^2 = 10^{-14}$ and $L = 5$.  }
\label{Fig:ContourEta}\end{figure}

To analyze the tradeoff between the increased SINR and reduction in the density of active transmissions as a result of increasing the radius of the guard-zone around receivers, we use the spectral efficiency density, given by $\eta = \rho(R_1) \log_2(1+\text{SINR})$, where  SINR is the SINR at the representative receiver which has a guard-zone of radius $R_1$ around it, and the density of active transmitters outside the guard-zone is $\rho(R_1)$. The CDF of the SINR is found by replacing  $X_T$ with $0$ in \eqref{Eqn:OffCenterIntens} which places a guard-zone around the representative receiver at the origin. The network-wide metric, $\eta$ assumes that the distances between interferers and their respective receivers are equal across the network. While this is a significant simplification, we note that in the absence of appropriate models for link-length distribution, this simplification is commonly used to optimize network-wide metrics such as in \cite{jindal2008bandwidth} and \cite{McKayGeneralized}.  In Fig. \ref{Fig:ContourEta}, we plotted the CDF of $\eta$,  $F_\eta (\eta; R_1)$ with $R_1$ varying from 0 to 10 and $\eta$ ranging from 0 to 0.1. Other parameters include  $\rho_p = 0.05$ ,  $r_T = 5$, $\alpha = 4$, $\sigma^2 = 10^{-14}$ and $L = 5$. Fig. \ref{Fig:ContourEta} shows that for a certain outage, there is an optimal radius of guard zone that maximizes the spectral efficiency density. Due to the complexity of finding an inverse function of $F_\eta (\eta; R_1)$ with respect to $R_1$, we found the optimal guard-zone radius can be evaluated numerically using standard zero-finding techniques. These plots indicate that a guard-zone could be useful in a wireless network even when an interference-mitigating multiantenna receiver is used since the optimal guard-zone radius is strictly positive in all the cases considered. Additionally, note
that the process of optimizing the guard-zone radius  via Monte Carlo simulation is very time consuming as a large number of different guard-zone radii
would have to be simulated and simulations of hard-core processes are relatively slow to begin with.  This illustrates that the  expressions we provide are useful to optimize system parameters when simulations would take prohibitively long to complete.  A similar idea can be used in networks which employ an ALOHA protocol over an existing doubly stochastic network model, e.g., a Poisson cluster process. The ALOHA probability of transmission can be maximized subject to a given outage probability constraint by numerically inverting the expression for the CDF in the same manner that was done to optimize the guard-zone radius. Optimizing this probability through Monte Carlo simulation would be significantly more challenging.


\section{Summary and Conclusions}\label{Sec:SummaryandConclusions}


In this paper, we develop a technique to compute the CDF of the SINR on a link with a multiantenna linear MMSE receiver in networks with co-channel
transmitters distributed according to doubly stochastic processes. Special cases of these processes include Poisson cluster processes,
processes with a single randomly located cluster and non-homogenous PPPs. This framework is applied to a variety of network models, including networks with deterministic and non-deterministic clusters strip networks, and hard-core networks.

Among others, these results enable us to quantify the benefits of multiple antennas in the center of a dense cluster of nodes. Moreover we find that CSMA-like protocols that place guard zones around receivers  can provide a spectral efficiency benefit, even when receivers use interference-mitigating receivers such as the MMSE receiver. Another interesting finding is that if the number of receiver antennas is increased linearly with the nominal density of nodes in doubly stochastic networks, then the SIR converges in distribution to a  random variable. Moreover, for non-homogenous Poisson networks the SIR converges in probability to a positive constant. This finding indicates that to the extent that the system assumptions hold (in particular the i.i.d. Rayleigh fading assumption), such networks are scalable by increasing the number of receiver antennas linearly with node density.

Thus, in addition to providing a framework to characterize the SINR of a multiantenna link in a broader class of network models than what's currently available, these results enable us to draw  conclusions regarding the design of spatially distributed networks as well.

\section{Acknowledgements}
We would like to thank the anonymous reviewers for the constructive comments.

\appendix

\subsection{Proof of Lemma \ref{Theorem:GenCDF}}\label{Sec:ProofOfGenCDF}
Established by equations (11) and (12) in \cite{ali2010performance}, the CDF of $\gamma$  can be expressed as
\begin{align*}
F_\gamma(\gamma) =1-\exp(-\sigma^2\gamma)\text{E}_n\left[\sum_{i=0}^{L-1}\sum_{k=0}^{\min(i,n)}\frac{n!(\sigma^2\gamma)^{i-k} }{k!(n-k)!(i-k)!} \text{E}_{p}\left[ \frac{p\gamma}{1+p\gamma}\right]^k   \text{E}_{p}\left[ \frac{1}{1+p\gamma}\right]^{i-k} \right]\,,
\end{align*}
where $\text{E}_x$ is the expectation with respect to the random variable $x$. Recall that $p = r^{-\alpha}$ and the interferer locations follow the PDF $f_{r,\theta}(r,\theta)$. Consequently, we have the following  expressions
\begin{align}
\text{E}_{p}\left[ \frac{ p\gamma}{1+p\gamma}\right]  &= \int_0^R\int_0^{2\pi} f_{r,\theta}(r,\theta)\frac{r^{-\alpha}\gamma}{1+r^{-\alpha}\gamma}d\theta dr\,,\label{E1}
\end{align}
\begin{align}
\text{E}_{p}\left[ \frac{1}{1+p\gamma}\right]  &=  \int_0^R\int_0^{2\pi} f_{r,\theta}(r,\theta) \frac{1}{1+r^{-\alpha}\gamma} d\theta dr\,.\label{E2}
\end{align}
As the number of interferers $n$ is a mean $\mu$ Poisson random variable, the CDF of $\gamma$ is:
\begin{align}
&  F_\gamma(\gamma) =1-\exp(-\sigma^2\gamma)\sum_{n=0}^{\infty}\sum_{i=0}^{L-1}\sum_{k=0}^{\min(i,N)}\frac{n!}{k!(n-k)!(i-k)!}(\sigma^2\gamma)^{i-k}    \frac{\mu^n}{n!}\exp(-\mu)   \cdot                                              \nonumber\\
&\;\;\;\left(\int_0^R\int_0^{2\pi} f_{r,\theta}(r,\theta)\frac{r^{-\alpha}\gamma}{1+r^{-\alpha}\gamma}d\theta dr \right)^k  \left(  \int_0^R\int_0^{2\pi} f_{r,\theta}(r,\theta) \frac{1}{1+r^{-\alpha}\gamma} d\theta dr \right)^{n-k}\,.
\end{align}
Applying a sequence of steps similar to that used in the proof of the main result for homogenous networks in \cite{ali2010performance} yields:
\begin{eqnarray}\label{longeqn}
 F_\gamma(\gamma) &=&1-\exp(-\sigma^2\gamma)\sum_{i=0}^{L-1}\sum_{k=0}^i \frac{(\sigma^2\gamma)^{i-k}}{k!(i-k)!} \left(\mu\int_0^R\int_0^{2\pi} f_{r,\theta}(r,\theta)\frac{r^{-\alpha}\gamma}{1+r^{-\alpha}\gamma}d\theta dr \right)^k  \cdot \nonumber\\
&&\exp\left(-\mu\int_0^R\int_0^{2\pi} f_{r,\theta}(r,\theta) \frac{r^{-\alpha}\gamma}{1+r^{-\alpha}\gamma} d\theta dr \right)\,.
\end{eqnarray}
Given the relationship between the PDF of the locations of the interferers and the intensity function in \eqref{PDFtoIndFunc}, we denote $\psi(\gamma)$ as:
\begin{equation}\label{Eqn:PsiDefInDeriv}
\psi(\gamma) = \lim_{R \rightarrow \infty}\int_0^R\int_0^{2\pi}\mu f_{r,\theta}(r,\theta)\frac{r^{-\alpha}\gamma}{1+r^{-\alpha}\gamma}d\theta dr = \int_0^\infty\int_0^{2\pi} \Lambda(r,\theta)r\frac{r^{-\alpha}\gamma}{1+r^{-\alpha}\gamma}d\theta dr\,.
\end{equation}
Substituting \eqref{Eqn:PsiDefInDeriv} into \eqref{longeqn}, applying the binomial theorem, the series expansion of the exponential function, and finally equation (6.5.13) of \cite{AbramovitzStegun} yields:
\begin{eqnarray}
 F_\gamma(\gamma) &=&1-\exp(-\sigma^2\gamma)\sum_{i=0}^{L-1}\sum_{k=0}^i \frac{(\sigma^2\gamma)^{i-k}}{k!(i-k)!} \psi^k(\gamma) \exp(\psi(\gamma))  \nonumber \\
&=&  1-\sum_{i=0}^{L-1}\frac{(\psi(\gamma)+\sigma^2\gamma)^i}{i!}\exp(-\psi(\gamma)-\sigma^2\gamma) =  1-\frac{\Gamma(L, \psi(\gamma)+\sigma^2 \gamma)}{\Gamma(L)}\,.
\end{eqnarray}
Differentiating the CDF yields the PDF of $\gamma$.


\subsection{Proof of Lemma \ref{Lemma:RegGammaLimit}} \label{Sec:ProofOfRegGammaLimit}

Consider the following random variables $U = \sum_{k = 0}^{L-1}
V_k$ and $\bar{U} = \frac{1}{L-1} \sum_{k = 1}^{L-1} V_k$,
where $V_k$ are i.i.d. Poisson random variables with mean $q$.
Note that
\begin{align}
Q(L, qL) &= \frac{\Gamma(L,qL)}{(L-1)!} =\Pr\left(U \leq L-1\right) = \Pr\left(\bar{U}+ V_0/(L-1) \leq 1\right). \label{Eqn:QtoCDF}
\end{align}
where the previous equality holds because $Q(L, qL)$ is the
probability that a Poisson random variable with mean $qL$ is
less than or equal to $L-1$. By the weak law of large numbers
as $L\to\infty$, both $\bar{U} \to q$ and $V_0/(L-1) \to 0$ in
probability implying that $\bar{U}+V_0/(L-1) \to q$ in
probability. The latter implies that
\begin{align}
\lim_{L\to\infty} \Pr\left(\bar{U}+\frac{V_0}{L-1}\leq u\right) = \begin{cases} 0, & \text{if $u\leq q$}
\\
1, &\text{if $u > q$}.
\end{cases}\label{Eqn:NormalizedPoissonSum}
\end{align}
Setting $u = 1$ in \eqref{Eqn:NormalizedPoissonSum} and
substituting \eqref{Eqn:QtoCDF} into the resulting equation
completes the proof

\subsection{Proof of Theorem \ref{Theorem:CDFConverge}}\label{Sec:ProofOfCDFConverge}

First, let us condition on a realization of the intensity function $\Lambda(r,\theta) = \lambda(r, \theta)$. The doubly stochastic network thus reduces to a non-homogenous PPP. Assuming that the noise is negligible, the CDF in \eqref{Eqn:GenCDF} from Lemma \ref{Theorem:GenCDF} can be expressed as:
\begin{align}
F_\gamma(\gamma | \Lambda = \lambda) = 1-\frac{\Gamma(L, \ell L\psi_c(\gamma; \lambda))}{\Gamma(L)}\,.
\end{align}
Let $q = \ell \psi_c(\gamma;\lambda)$ in Lemma
\ref{Lemma:RegGammaLimit}. Then, we have:
\begin{align}\label{Eqn:CDFConvergeCond}
 \lim_{L \to \infty}F_\gamma(\gamma;\lambda) = 1-\lim_{L\to\infty}\frac{\Gamma(L, \ell L\psi_c(\gamma))}{\Gamma(L)} = \phi(\gamma; \lambda)=\begin{cases} 0,  & \text{if $\gamma \le \psi_c^{-1}\left(\frac{1}{\ell };\lambda\right) $}      \\     1, &\text{if $\gamma > \psi_c^{-1}\left(\frac{1}{\ell };\lambda\right)$}.
\end{cases}
\end{align}
Thus, conditioned on particular realization of $\Lambda(r, \theta)$ or if the
interferers form a  non-homogenous Poisson process (i.e. deterministic $\Lambda(r,\theta)$), $\gamma$ converges in distribution to a constant, implying convergence in probability as well.
Removing the conditioning in \eqref{Eqn:CDFConvergeCond} , we have
\begin{align}\label{Eqn:CDFConverge}
 \lim_{L \to \infty}F_\gamma(\gamma)= \lim_{L \to \infty}\mathbf{E}_{\Lambda}\left[F_\gamma(\gamma;\lambda)\right]= \mathbf{E}_{\Lambda}\left[\lim_{L \to \infty} F_\gamma(\gamma;\lambda)\right] = \mathbf{E}_\Lambda[\phi(\gamma; \lambda)]\,,
\end{align}
where  the exchange of the expectation and limit operations follows from the bounded convergence theorem (see e.g., \cite{Karr}). Note that unlike the case of non-homogenous Poisson process of interferers, in the case of a general doubly stochastic process of interferers, $\gamma$ is only guaranteed to converge in distribution.

\subsection{Proof of Lemma \ref{Lemma:Campbell}} \label{Sec:Campbell}

It is shown in Section 3.2 of \cite{Kingman} that given any positive measurable function $\zeta(.)$,
\begin{align}\label{Eqn:ExParents}
\mathbf{E}_{\Pi}[e^{\Theta \Xi}] = \exp{\left\{ \int_{\mathbb{R}^2}  \left( e^{\Theta \zeta(u)}-1  \right)  \lambda(u) du\right\} }\,,
\end{align}
for any negative and real $\Theta$. Taking the $k$th  derivative with respect to $\Theta$, \eqref{Eqn:ExParents} becomes
\begin{eqnarray}
&&\mathbf{E} _{\Pi}\left[\Xi^k e^{\Theta \Xi}\right]= \frac{d^k}{d\Theta^k} \exp{\left\{ \int_{\mathbb{R}^2} \left( e^{\Theta \zeta (u)}-1  \right) \lambda(u)  du\right\} } \nonumber \\
&=& \sum_{(m_1,...,m_k) \in \mathcal{M}_k}  \frac{k!\exp \left\{ \int_{\mathbb{R}^2}  \left( e^{\Theta \zeta(u)}-1  \right) \lambda(u)  du \right\}}{m_1! 1!^{m_1}m_2! 2!^{m_2}...m_k! k!^{m_k}} \prod_{j=1}^{k}\left( \int_{\mathbb{R}^2}  \zeta^j(z) e^{\Theta \zeta (z)} \lambda(z) dz \right)^{m_j}.\label{Eqn:ModifiedCampbell}
\end{eqnarray}
The last step follows from Fa\`{a} di Bruno's formula \cite{successivederivatives}:
\begin{eqnarray*}
\frac{d^k}{d\Theta^k}f(g(\Theta)) = \sum_{(m_1,...,m_k) \in \mathcal{M}_k} \frac{k!}{m_1! 1!^{m_1}m_2! 2!^{m_2}...m_k! k!^{m_k}}f^{(m_1+...+m_k)}\left(g(\Theta) \right) \cdot \prod_{j=1}^{k}\left(g^{(j)}(\Theta) \right)^{m_j}\,,
\end{eqnarray*}
where $f(.)$ and $g(.)$ are arbitrary measurable functions.
Evaluating \eqref{Eqn:ModifiedCampbell} at $\Theta=-1$ yields
\eqref{Eqn:Lemma2}.

\subsection{General Power Law Term}\label{Sec:HypergeometricFunction}
\begin{lemma} \label{Lemma:HypergeometricFunction}
Given the assumption that $\alpha>2$,
\begin{align} \label{Eqn:HypergeometricFunction}
\int r^{\kappa+1} \frac{\gamma}{r^\alpha + \gamma}dr \bigg{|}_{r=R'} = \begin{cases} 0, & \text{if $R' = 0$}\\
\frac{R'^{\kappa+2}}{\kappa+2} {_2F_1} \left(1,\frac{\kappa+2}{\alpha};\frac{\kappa+2+\alpha}{\alpha};-\frac{R'^\alpha}{\gamma}\right) , &\text{if $0<R'<\infty$}\\
\frac{\pi}{\alpha}\gamma^{(\kappa+2)/\alpha}\csc\left(\pi \frac{\kappa+2}{\alpha} \right), &\text{if $R'=\infty$}
.
\end{cases}
\end{align}
\end{lemma}

{\it Proof:} Using symbolic integration software, we can directly evaluate the integral for $0< R'< \infty$.
Applying Euler's hypergeometric transformation \cite{AbramovitzStegun} to the resulting expression, taking $R'\to\infty$ and applying the reflection formula for hypergeometric functions (see e.g., \cite{AbramovitzStegun}, yields the third expression in the case statement in \eqref{Eqn:HypergeometricFunction}

\bibliographystyle{IEEEbib}
\bibliography{IEEEabrv,main}

\end{document}